\documentclass[aps,prd,showpacs,twocolumn]{revtex4}

\usepackage{epsfig}
\usepackage{longtable}
\usepackage{graphicx}

\begin{document}

\title{Exploring Higgs Triplet Models via Vector Boson Scattering at the LHC}
\author{ Stephen Godfrey$^{1,2}$\footnote{Email: godfrey@physics.carleton.ca} 
and Ken Moats$^{1}$\footnote{Email: kmoats@physics.carleton.ca}}
\affiliation{
$^1$Ottawa-Carleton Institute for Physics, 
Department of Physics, Carleton University, Ottawa, Canada K1S 5B6\\
$^2$TRIUMF, 4004 Wesbrook Mall, Vancouver BC Canada V6T2A3 
}

\date{\today}

\begin{abstract}
We present the results of a study of Higgs triplet boson production arising in the Littlest
Higgs, Left-Right Symmetric, and Georgi-Machacek models in the $W^\pm W^\pm$,
$W^\pm Z$, $W^+ W^-$, and $Z Z$ channels at the LHC.  
We focus on the ``gold-plated'' purely leptonic decay modes and consider the
irreducible electroweak, QCD, and $t$-quark backgrounds, applying a combination of forward-jet-tagging, 
central-jet-vetoing, and stringent leptonic cuts to suppress the backgrounds.  We find 
that, given the constraints on the triplet vacuum expectation value (vev), considerable
luminosity is required to observe Higgs triplet bosons in vector boson scattering.  
Observing a Higgs triplet at the LHC is most promising in the Georgi-Machacek model due to 
a weaker constraint on the triplet vev.  In this model, we find that a Higgs triplet boson 
with a mass of 1.0 (1.5) TeV can be observed at the LHC with an integrated luminosity as low 
as 41 (119) fb$^{-1}$ in the $W^\pm W^\pm$ channel and as low as 171 (474) fb$^{-1}$ in the $W^\pm Z$ channel.
Observation of Higgs triplet bosons in these channels would help identify the underlying theory.

\end{abstract}
\pacs{12.60.Fr, 14.80.Fd, 12.15.Ji}

\maketitle

\section{Introduction}
A primary motivation for the CERN Large Hadron Collider is to understand the mechanism of
electroweak symmetry breaking (EWSB) \cite{Morrissey:2009tf,Accomando:2006ga,Nath:2010zj}. 
Related to understanding EWSB is the need for physics beyond the Standard Model (SM)
to resolve the hierarchy and fine tuning problems between the electroweak scale
and the Planck scale.  
Numerous solutions, such as Supersymmetry, Dynamical Symmetry 
Breaking, and Extra Dimensions, have been proposed \cite{Morrissey:2009tf,Accomando:2006ga,Nath:2010zj}.  
An interesting class of models known
as ``Little Higgs'' models \cite{ArkaniHamed:2002qy} has a very rich phenomenology, including new 
heavy gauge bosons, new heavy quarks, and an expanded scalar sector.  Much of this has been
studied elsewhere 
\cite{Burdman:2002ns,Han:2003wu,Han:2005ru,Perelstein:2005ka,Azuelos:2004dm,Hubisz:2004ft} 
with, for the most part, the exception of the scalar sector of the theory. 
Would-be Goldstone multiplets are a fundamental ingredient of these theories, and 
the existence of scalar triplets might be a useful signature for this class of models.

It has been pointed out that doubly-charged members of the Higgs triplet ($\Phi^{\pm\pm}$) might
serve as a good signal for these models if they are kinematically accessible at future 
colliders \cite{Han:2003wu}. Han {\it et al.} \cite{Han:2003wu} studied 
the effects of resonant contributions 
from $\Phi^{++}\to W^+W^+$  in an idealized calculation of longitudinal $W^+ W^+$ scattering, 
$W^+_L W^+_L \to W^+_L W^+_L$, in the Littlest Higgs model. A more detailed analysis by 
Azuelos {\it et al.} \cite{Azuelos:2004dm} found that a $\Phi^{++}$ could be
observed up to 2~TeV with the right range of parameters.  However, the Littlest Higgs model also 
predicts neutral ($\Phi^0$) and singly-charged ($\Phi^\pm$) members of the Higgs triplet with 
$W^+W^- \Phi^0$, $ZZ\Phi^0$, and $W^+ Z \Phi^-$ interactions.  If heavy scalars were 
observed, it would be necessary to map out their properties to construct the underlying Lagrangian.
For example,  the $W^+ Z \Phi^-$ vertex is unique to Higgs triplet representations and does not
exist in Two Higgs Doublet Models (2HDM), so the observation of this interaction would provide evidence 
for Higgs triplets \cite{Asakawa:2006gm}.

The observation of a Higgs triplet state would not, however, uniquely identify the model of 
origin \cite{Asakawa:2006gm}.  
In addition to the Littlest Higgs model, scalar triplets are also included in the Georgi-Machacek 
model \cite{Georgi:1985} and Left-Right Symmetric models \cite{Mohapatra:1975, Senjanovic:1975rk}.
The so-called 3-3-1 models \cite{Pisano:1991ee,Frampton:1992wt} 
are another class of models that include Higgs triplets
\cite{CiezaMontalvo:2006zt}.  However, because 3-3-1 models are similar to variations of Little Higgs
models, we will not consider them separately.  To distinguish between models, one 
therefore needs to study the properties of newly discovered scalar resonances.
In this paper, we study vector boson scattering to determine whether it can be used to distinguish 
between models that include Higgs triplet bosons.  
Vector boson pair production was also considered as a signature of Higgless Models 
\cite{Csaki:2003dt,Csaki:2003zu,Belyaev:2009ve,He:2007ge,Alves:2009aa,Birkedal:2004au}
and other possibilities for new physics \cite{Cheung:2008zh,Dorsner:2007fy}
but we do not include these scenarios in this paper.

Higgs bosons can in principle couple to fermions and, in fact, a same sign dilepton final state would
almost certainly be a cleaner signal for a doubly-charged Higgs boson than vector boson final states.  
Although this lepton number violating decay, $\Phi^{\pm \pm} \to l^{\pm}l^{\pm}$, does not occur in the version 
of the Littlest Higgs model that we consider, it can occur in certain variations of the Littlest Higgs 
model \cite{Han:2005nk}, as well as in the Georgi-Machacek \cite{Gunion:1989ci} and Left-Right Symmetric 
models \cite{Huitu:1996su, Gunion:1989in, Azuelos:2005uc}.  It is therefore worth commenting on 
the relative strengths of the dilepton and $VV$ channels. Neutrino masses are an important constraint on 
the allowed parameter range, which in turn leads to different decay scenarios for the Higgs triplet 
states \cite{Han:2005nk}.  In one case, the small
neutrino mass can be driven by a tiny triplet vacuum expectation value (vev), but with neutrino Yukawa couplings 
${\cal O}(1)$, while in the opposite scenario it is the Yukawa couplings that are tiny to accomodate the
observed neutrino masses.  It is the latter case which can lead to a triplet vev that is large enough to 
give rise to the signal we are studying.  

In this paper, we report on a study of heavy scalar triplet production via vector boson scattering 
in the $W^\pm W^\pm$, $W^\pm Z$, $W^+W^-$, and $Z Z$ channels. 
The models we consider are the 
Littlest Higgs Model, the Georgi-Machacek Model and the Left-Right Symmetric Model.
We focus on the purely leptonic decays of the vector bosons, which have the advantage
of lower SM QCD backgrounds.  The final states we consider are therefore 
$W^\pm W^\pm\to l^\pm \nu  l^\pm \nu$,  $W^\pm Z \to l^\pm \nu l^+l^-$,
$W^+W^- \to l^+ \nu l^- \bar{\nu}$ and $ZZ\to l^+l^- l^+l^-$.  
These processes have been extensively studied in the context
of strongly interacting weak sector models 
\cite{Barger:1990py,Bagger:1993zf,Bagger:1995mk,Kilgore:1996pj} and previous studies
of doubly-charged Higgs boson production at the LHC in the Littlest Higgs and Left-Right Symmetric models
were given by Azuelos {\it et al.} \cite{Azuelos:2004dm, Azuelos:2005uc}.

In the following section, we give a brief introduction to the models we study in this paper, 
focusing on properties most relevant to our analysis.  This is followed  in Section III
by a description of
calculational details.  In Section IV we describe possible backgrounds and the kinematic cuts 
that can be used to reduce them.  Our results are given in Section V followed by 
a brief summary of our conclusions in Section VI.

\section{Models}

In this section, we give a brief overview of the Littlest Higgs, Georgi-Machacek, and Left-Right
Symmetric models. It is not
intended to be an extensive review but only gives details used in our analysis.
The Feynman rules needed for our calculations are summarized in Table~\ref{tab:feynman}.

\begin{table}[b]
\begin{center}
\caption{Feynman Rules for the interactions of scalar and vector bosons in the Littlest Higgs \cite{Han:2003wu}, 
Georgi-Machacek \cite{Gunion:1989ci} and Left-Right Symmetric models \cite{Gunion:1989in}.} 
\label{tab:feynman}
\begin{ruledtabular}
\begin{tabular}{ll}
\multicolumn{2}{c}{Littlest Higgs}   \\
\hline
$W^+_{\mu} W^-_{\nu} h$ & 	
	$\frac{i}{2} g^2 v \left( 1 
	- \frac{v^2}{3f^2}- \frac{1}{2} (c^2-s^2)^2 \frac{v^2}{f^2}
 - \frac{1}{2} s_0^2  \right.$ \\ & $ \qquad\qquad\qquad\qquad\qquad\qquad\quad \left. - 2 \sqrt{2} s_0 \frac{v^{\prime}}{v}\right) g_{\mu\nu}  $  \\

$Z_{\mu} Z_{\nu} h$ &	
	$\frac{i}{2} \frac{g^2}{c^2_W} v 
	\left( 1 - \frac{v^2}{3f^2} - \frac{1}{2} s_0^2 + 4 \sqrt{2} s_0 \frac{v^{\prime}}{v} \right.$ \\ & $ \qquad\quad \left. -\frac{1}{2} \left( (c^2-s^2)^2 + 5 (c^{\prime 2}-s^{\prime 2})^2 \right) \frac{v^2}{f^2}\right) g_{\mu\nu}$ \\

$W_{\mu}^+ W_{\nu}^- \Phi^0$ &	 $-\frac{i}{2} g^2 ( s_0 v - 2 \sqrt{2} v^{\prime} ) g_{\mu\nu} \simeq 0$  		\\

$Z_{\mu} Z_{\nu} \Phi^0$ &	$-\frac{i}{2} \frac{g^2}{c^2_W} ( s_0 v - 4 \sqrt{2} v^{\prime} )g_{\mu\nu} \simeq \sqrt{2} i \frac{g^2}{c^2_W}  v^{\prime} g_{\mu\nu}$	\\

$W_{\mu}^+ Z_{\nu} \Phi^-$ &	 $-i \frac{g^2}{c_W} v^{\prime} g_{\mu\nu}$		\\

$W_{\mu}^+ h \Phi^-$ &	 $-i \frac{g}{2} ( \sqrt{2} s_0  - s_+ ) (p_1-p_2)_\mu \simeq -i g \frac{v'}{v} (p_1-p_2)_\mu $	\\

$W_{\mu}^+ W_{\nu}^+ \Phi^{--}$ &	 $2i g^2 v^{\prime} g_{\mu\nu}$	\\

\hline
\multicolumn{2}{c}{Georgi-Machacek}   \\
\hline
$W^+_{\mu} W^-_{\nu} H_1^{0} $ & 	$\frac{i}{2} g^2 v \left( 1 - 8 \frac{v'^2}{v^2}\right)g_{\mu\nu} $  \\

$Z_{\mu} Z_{\nu} H_1^{0} $ &	$\frac{i}{2} \frac{g^2}{c^2_W} v \left( 1 - 8\frac{v'^2}{v^2}\right) g_{\mu\nu} $ \\

$W_{\mu}^+ W_{\nu}^- H_5^0$ & $ \sqrt{\frac{2}{3}} i g^2 v' g_{\mu\nu}$  		\\

$Z_{\mu} Z_{\nu} H_5^0$ &	$- 2 \sqrt{\frac{2}{3}} i \frac{g^2}{c_W^2} v' g_{\mu\nu}$  	\\

$W_{\mu}^+ Z_{\nu} H_5^-$ & $- \sqrt{2} i  \frac{g^2}{c_W} v' g_{\mu\nu}$ 		\\

$W_{\mu}^+ H_1^{0} H_5^-$ & $0$ 		\\

$W_{\mu}^+ W_{\nu}^+ H_5^{--}$ &  $2 i g^2 v' g_{\mu\nu}$ 	\\

$W_{\mu}^+ W_{\nu}^- H_1^{0 \prime}$ & $ \frac{4}{\sqrt{3}} i g^2 v' g_{\mu\nu}$  		\\

$Z_{\mu} Z_{\nu} H_1^{0 \prime}$ &$ \frac{4}{\sqrt{3}} i  \frac{g^2}{c_W^2} v' g_{\mu\nu}$    	\\

\hline
\multicolumn{2}{c}{Left-Right Symmetric}   \\
\hline
$W^+_{\mu} W^-_{\nu} h $ & 	$\frac{i}{2} g^2 v g_{\mu\nu} $  \\

$Z_{\mu} Z_{\nu} h $ &	$\frac{i}{2} \frac{g^2}{c^2_W} v g_{\mu\nu} $ \\

$W_{\mu}^+ W_{\nu}^- \Delta_L^0$ & $ \sqrt{2} i g^2 v' g_{\mu\nu}$  	\\

$Z_{\mu} Z_{\nu} \Delta_L^0$ &	 $ 2\sqrt{2} i \frac{g^2}{c_W^2} v' g_{\mu\nu}$ 		\\

$W_{\mu}^+ Z_{\nu} \Delta_L^-$ &  $ - i \frac{g^2}{c_W} v' g_{\mu\nu}$	\\

$W_{\mu}^+ h \Delta_L^0$ &	 $ 0$ 		\\

$W_{\mu}^+ W_{\nu}^+ \Delta_L^{--}$ & $- 2 i g^2 v' g_{\mu\nu}$	\\

\end{tabular}
\end{ruledtabular}
\end{center}
\end{table}

\subsection{The Littlest Higgs Model} \label{sec:littlest_higgs}

Little Higgs models \cite{ArkaniHamed:2002qy,ArkaniHamed:2001nc,Schmaltz:2005ky} 
are a class of models that attempt to resolve the hierarchy and fine-tuning problems between 
the electroweak scale and the Planck scale.  There are many variations of Little Higgs models 
(for recent reviews, see  \cite{Han:2003wu,Han:2005ru,Schmaltz:2005ky,Perelstein:2007}).  
In this paper, we consider the Littlest Higgs Model \cite{ArkaniHamed:2002qy}, 
which introduces new particles at a scale  $f \sim 1$ TeV in addition to the particles of the 
Standard Model.  The couplings of these particles to the SM Higgs boson are such that the quadratic 
divergences 
to $M_{h}^{2}$ introduced by the SM loops are cancelled by the 
quadratic divergences introduced by the new TeV scale particles at one-loop level.  As a result, 
the SM Higgs boson remains light and free from one-loop quadratic sensitivity up 
to a cutoff scale, $\Lambda_{S} = 4 \pi f \sim 10$ TeV.

The scalar field content of the Littlest Higgs model consists of a doublet, $h$, and a triplet, 
$\phi$, under the
unbroken $SU(2)_L\times U(1)_Y$ Standard Model gauge group:
	\begin{equation}
	h = \left( \begin{array}{c}
			h^{+}  \\
			h^0
		\end{array} \right), \qquad
	\phi = \left( \begin{array}{cc}
			\phi^{++} & \phi^+ / \sqrt{2} \\
			\phi^+ / \sqrt{2} & \phi^0
		\end{array} \right).
	\end{equation}
Electroweak symmetry breaking results in a vev for the neutral components of
both the doublet and the triplet fields: $\langle h^{0}\rangle = v/\sqrt{2}$ 
and $\langle i\phi^{0}\rangle = v'$.  The doublet and triplet states acquire masses, 
the latter of which is given to leading order by \cite{Han:2003wu}
\begin{equation} \label{higgs_masses}
 M^2_\Phi=\frac{2M_h^2 f^2}{v^2}
\frac{1}{\left[1- (4v^{\prime} f / v^2)^2 \right]}.
\label{MPHI}
\end{equation}
By demanding that $M_{\Phi}^{2} > 0$, we arrive at a 
relation between the doublet and triplet vevs \cite{ArkaniHamed:2002qy,Han:2003wu}:
\begin{equation}
{v'\over v} < { v\over 4f}.
\label{vprime}
\end{equation}
For $f = 2 \;\rm TeV$ and $v \approx 246 \;\rm GeV$, this gives an upper bound
\begin{equation}
v' < 8 \hbox{ GeV}.
\label{upper_limit}
\end{equation}
The value of $v'$ has been further constrained by electroweak data to
1 GeV $\lesssim v' \lesssim 4$ GeV for $f = 2$ TeV \cite{Chen:2004}.

\subsection{The Georgi-Machacek Model}  \label{sec:Georgi-Machacek}

The next model we consider is the Georgi-Machacek model \cite{Georgi:1985}, 
in which the scalar fields take the form 
\begin{equation}
	\phi = \left( \begin{array}{c}
			\phi^+ \\
			\phi^0
		\end{array} \right), \qquad
	\chi = \left( \begin{array}{ccc}
			\chi^{0} 	& \xi^{+}	& \chi^{++} \\
			\chi^{-} 	& \xi^{0}	& \chi^{+} \\
			\chi^{--} 	& \xi^{-}	& \chi^{0^{*}}
		\end{array} \right),
	\end{equation}
where $\phi$ is a complex doublet, $\xi$ is a real triplet and $\chi$ is a complex triplet.
Electroweak symmetry breaking occurs when the neutral components of these fields develop a non-zero vev,
 given by $\langle \phi^{0} \rangle = a/\sqrt{2}$ and
$\langle \chi^{0} \rangle = \langle \xi^{0} \rangle = v'$.  It is convenient 
to introduce the notation
\begin{eqnarray}\label{georgi_vev1}
	v^2 &\equiv& a^2 + 8v'^2, \\
	c_H^2 &\equiv& \frac{a^2}{v^2}, \quad
	s_H^2 \equiv \frac{8 v'^2}{v^2},\label{georgi_vev2}
\end{eqnarray}
as the $W^{\pm}$ and $Z$ bosons masses are given by
\begin{equation}\label{georgi_gauge_masses}
M_W^2 = M_Z^2 \cos^2\theta_W = \frac{1}{4} g^2 (a^2 + 8v'^2) = \frac{1}{4} g^2 v^2.
\end{equation}
We substituted the relations given by Eqns.~\ref{georgi_vev1}, \ref{georgi_vev2}
and \ref{georgi_gauge_masses} into
the Feynman rules given by Ref.~\cite{Gunion:1989ci} so that the Feynman rules in 
Table~\ref{tab:feynman} use a common notation. 

The physical scalars in this model can be classified according to their transformation 
properties under the custodial SU(2) symmetry, which also guarantees that $\rho = 1$ at tree-level.  
One finds a five-plet $(H_5^{\pm \pm}$, 
$H_5^{\pm}$, $H_5^{0})$, a three-plet $(H_3^{\pm}$, $H_3^{0})$, and two singlets, $H_1^0$ 
(which is identified as the SM Higgs boson) and $H_1^{0 \prime}$.  In this paper, we are interested 
in the production of the $H_5$ 
triplet as these particles couple to the $W^{\pm}$ and $Z$ bosons, whereas the $H_3$ 
triplet does not.

From Equation \ref{georgi_gauge_masses}, we see that $v = \sqrt{a^2 + 8v'^2} \approx 246$ GeV, 
but the ratio $v'/a$ is not fixed.  
The strongest experimental bound on this ratio comes from the $Zb\overline{b}$ coupling measurements, 
which give, at 95\% C.L., $\tan \theta_H \equiv s_H / c_H \lesssim 0.5, 1 $ and 1.7 
for $M_{H_3} = 0.1, 0.5$ and 1 TeV, respectively \cite{Haber:2000}. A value of $\tan\theta_H=0.5$
corresponds to a triplet vev of $v'=39$~GeV, which we use in all our calculations.

\subsection{The Left-Right Symmetric Model} \label{sec:LR_symmetric}

The scalar sector of the $SU(2)_L \times SU(2)_R \times U(1)_{B-L}$ Left-Right Symmetric 
model \cite{Mohapatra:1975,Senjanovic:1975rk} is given by \cite{Gunion:1989in,Deshpande:1990ip}
\begin{eqnarray}
	\phi &=& \left( \begin{array}{cc}
				\phi_1^0 & \phi_1^+ \\
				\phi_2^-  & \phi_2^0
			\end{array} \right), \\
	\Delta_L &=& \left( \begin{array}{cc}
				\Delta_L^+ / \sqrt{2} 	 & \Delta_L^{++} \\
				\Delta_L^0 		 & - \Delta_L^+ / \sqrt{2}
			\end{array} \right), \\
	\Delta_R &=& \left( \begin{array}{cc}
				\Delta_R^+ / \sqrt{2} 	& \Delta_R^{++} \\
				\Delta_R^0 		& - \Delta_R^+ / \sqrt{2}
			\end{array} \right),
\end{eqnarray}
where left-right symmetry requires that the Lagrangian of the theory is invariant under $\Delta_L \leftrightarrow \Delta_R$ and $\phi \leftrightarrow \phi^\dagger$.  The vacuum expectation values of the doublet and triplet are given by
\begin{eqnarray}
	\langle \phi \rangle &=& \frac{1}{\sqrt{2}}	\left( \begin{array}{cc}
										\kappa_1 & 0 \\
										0  & \kappa_2
									\end{array} \right), \\
	\langle \Delta_{L,R} \rangle &=& \frac{1}{\sqrt{2}}	\left( \begin{array}{cc}
												0 & 0 \\
												v_{L,R} & 0	
												\label{left_triplet}																\end{array} \right).
\end{eqnarray}
The triplet vevs $(v_L, v_R)$ break 
$SU(2)_L \times SU(2)_R \times U(1)_{B-L} \to SU(2)_L \times U(1)_Y$, 
while the doublet, $\phi$, breaks $SU(2)_L \times U(1)_Y \to U(1)_{EM}$.  

In this paper we are interested in the production of the left triplet, $\Delta_{L}$, as it 
couples to the Standard Model $W^{\pm}$ and $Z$ bosons.  The left triplet vev, $v_L$, is 
constrained by the $\rho$ parameter, given by
\cite{Gunion:1989in,Huitu:1996su}:
\begin{equation}
\rho = \frac{M_W^2}{\cos^2 \theta_W M_Z^2} \simeq \frac{1+2v_L^2 /v^2}{1+4v_L^2 /v^2}.
\end{equation}
The experimental result $\rho = 1.0004 ^{+0.0027}_{-0.0007}$ at $2\sigma$ \cite{Amsler:2008zzb} implies 
that $v_L \lesssim 3$ GeV, a small value compared with 
$v \equiv \sqrt{\kappa_1^2 + \kappa_2^2} \approx 246$ GeV.  Note that due to the factor 
of $1/\sqrt{2}$ in Equation \ref{left_triplet}, 
we define the triplet vev in this model as 
$v' \equiv v_L/\sqrt{2}$ to be consistent with the conventions of the other models under study.  
Using this notation, the upper limit on the triplet vev becomes  $v' \lesssim 2$ GeV.

\section{Calculations}

\begin{figure}[b]
  \begin{center}
 \includegraphics[width=2.0in]{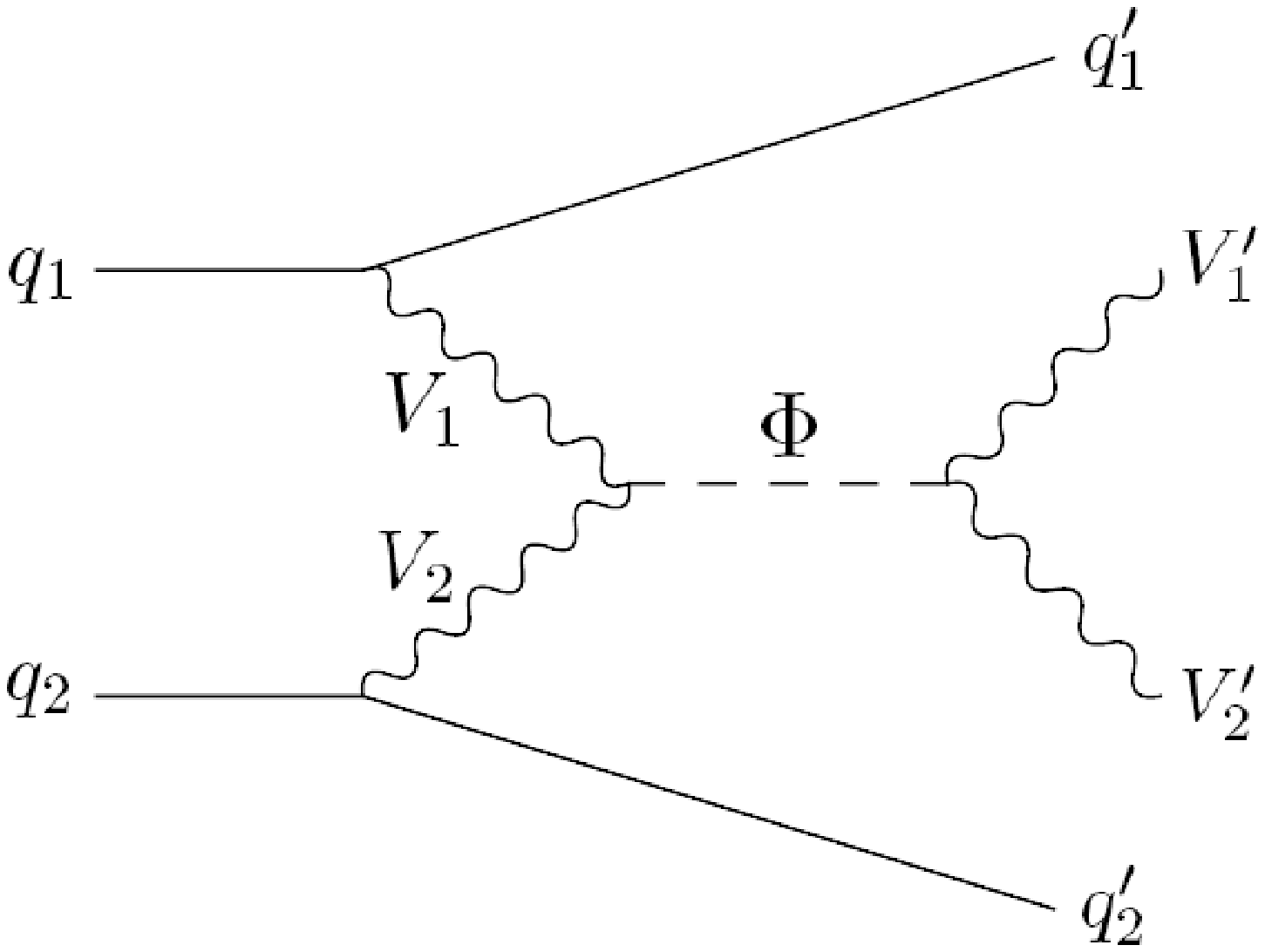}
 \caption{Higgs triplet production via vector boson scattering at the LHC.  The incoming quarks 
from the colliding protons radiate vector bosons, which interact to produce a Higgs triplet 
as an s-channel resonance.  Two vector bosons emerge in the final state, along with two forward 
spectator quark jets. }
   \label{fig:VBF_diagram}
   \end{center}
   \end{figure}

The production mechanism for scalar triplet bosons via vector boson scattering at the LHC is
illustrated in Fig.~\ref{fig:VBF_diagram}.
The incoming quarks from the 
colliding protons radiate vector bosons ($V = W^\pm, Z$), which interact  to produce the Higgs triplet.
The Higgs triplet then decays to vector bosons, accompanied by two spectator quark jets that emerge in 
the forward region of the detector.  The
forward jets are the hallmark of the vector boson fusion process and are an 
important tool to reduce backgrounds. We study the production of all members of the 
scalar triplet: $\Phi^{\pm \pm}$ via $W^\pm W^\pm$ scattering,  $\Phi^\pm$ via  $W^\pm Z$ scattering, 
and $\Phi^0$ via $W^+ W^-$ and $Z Z$ scattering. 

To calculate the $pp\to \Phi + jj \to VV +jj$ cross sections and kinematic distributions, we used the 
MadGraph software package \cite{Alwall:2007st}. MadGraph generates tree-level matrix elements 
and then generates events using Monte Carlo phase space integration.  We then analyzed the
generated events, including implementing kinematic cuts.  We show the resulting 
kinematic distributions in figures and the cross sections in tables.

We implemented the relevant scalar sector 
Feynman rules for the Littlest Higgs model \cite{Han:2003wu}, the Georgi-Machacek model \cite{Gunion:1989ci} 
and the Left-Right Symmetric model \cite{Gunion:1989in} in MadGraph (see Table \ref{tab:feynman}).  
Note that in the Littlest Higgs model, the scalar sector mixing angles are given to leading order by
$s_0 = \sqrt{2} s_+ \simeq 2\sqrt{2} v^\prime/v$ \cite{Han:2003wu}.  This results in a value of zero 
for the $W^+ W^- \Phi^0$ vertex so we did not study the $W^+ W^-$ scattering process in the 
Littlest Higgs model.  In the $WWh$ and $ZZh$ vertices of the Littlest Higgs model, we did not include 
the $\mathcal{O}(v^2/f^2)$ terms in our calculations since their contributions are negligible for our 
purposes, regardless of the values of the mixing angles ($s, c, s^\prime, c^\prime$)
in the gauge sector of this model. 

It is interesting to note some of the similarities and differences in the Feynman rules of each model.  
For example, the $W^+ W^+ \Phi^{--}$ vertex has the same form in all three models.  This is due to the 
fact that the doubly-charged member of the triplet has the same EW quantum numbers in each model.  
However, because the $H_5^{\pm}$ of the Georgi-Machacek model is a mixture of the 
$\chi^{\pm}$ and $\xi^{\pm}$ states, which have $(T_3, Y) = (0, \pm2)$ and $(\pm1, 0)$ respectively, 
this model does not have the same $W^+ Z \Phi^{-}$ vertex as in the Littlest Higgs and Left-Right Symmetric 
models. 

In principle, one could also use vector boson scattering to study the production of the 
$H_1^{0 \prime}$ singlet in the Georgi-Machacek model.  From the Feynman rules in Table \ref{tab:feynman}, 
we see that it couples more strongly to the $W$ and $Z$ bosons than the $H_5$ triplet does.  Therefore, 
depending on its mass, the $H_1^{0 \prime}$ may be more easily observable in the $W^+W^-$ and $ZZ$ channels.  
Production of the $H_1^{0 \prime}$ is also a potential background to the $H_5^0$ production process, which 
we are interested in studying.
However, because we are focusing on members of the Higgs triplet, 
for simplicity, we do not include $H_1^{0 \prime}$ production in our analysis.

Hadronic decays of the vector bosons have the largest branching ratios.  However, the QCD backgrounds to these 
final states are very large and difficult to disentangle from the signal.  In contrast, the purely leptonic 
decays, 
with $W^{\pm} \rightarrow l^{\pm} \nu$ 
and $Z \rightarrow l^{+} l^{-} \; (l = e, \mu)$, provide
a much cleaner signal than hadronic decays, at the cost of a smaller branching ratio.  These purely leptonic
 final states are often referred to as ``gold-plated" modes.  
A final possibility is to consider semileptonic 
final states, sometimes referred to as ``silver-plated'' modes, which are intermediate in terms of branching 
ratios and clean final states.  
As expertise for reconstructing
vector bosons in hadronic modes improves, these final states could prove to be useful for studying
vector boson scattering.
However, in this paper we focus on the ``gold-plated'' leptonic final states.

MadGraph includes many diagrams with the same leptonic final state as the process in which we are interested.  
Therefore, having MadGraph calculate the full process including decays to final state leptons includes a very
 large number of non-resonant diagrams and takes considerable computing time.  
Instead, we used MadGraph to calculate
the $pp\to \Phi + jj \to VV +jj$ processes and subsequently decayed the vector bosons using 
the DECAY package of MadGraph.  We compared cross sections and distributions 
using both approaches and they agreed within the statistical uncertainties of the Monte Carlo.

As a final check of our results, we calculated cross sections using 
the Effective Vector Boson
Approximation \cite{Dawson:1985} and the Goldstone Equivalence Theorem \cite{Cornwall:1974}, and found that the two approaches were in 
reasonable agreement \cite{moats}.  MadGraph includes a more complete 
set of Feynman diagrams that contributes to the final states we are studying and, more 
importantly, includes the spectator jets which are a crucial ingredient for 
reducing backgrounds.

\section{Backgrounds} \label{sec:backgrounds}

At the high energies associated with TeV scale Higgs boson production at the LHC, the cross section 
for vector boson scattering is dominated by scattering of longitudinally polarized gauge bosons, 
$ V_{L}V_{L}\to \Phi \to V_{L}V_{L}$, where $V = W^{\pm}, Z$.  Therefore, the signal process in which we are 
interested is
\begin{equation}
pp \rightarrow V_{L}V_{L}jj,
\end{equation}
 with both vector bosons decaying leptonically.  
In general, the experimental signature consists of high-$p_T$ leptons in the central region of the detector, 
two high-$p_T$ forward jets and low jet activity in the central region. 
We consider three types of backgrounds to this signal following 
the treatment of
\cite{Barger:1990py,Bagger:1993zf,Bagger:1995mk}.

The first background is the irreducible electroweak (EW) background.  It
is similar to the signal process except that
at least one of the vector bosons is transversely polarized: 
\begin{eqnarray}
pp & \to & V_{T}V_{L}jj, \\
pp & \to & V_{T}V_{T}jj.
\end{eqnarray}
This process is a background in 
the sense that its cross section is 
essentially insensitive to the new TeV scale physics that we wish to study.  It
is irreducible in the sense that the final state is virtually identical to the signal process, 
differing only by the polarization of the vector bosons, making this background difficult to reduce using 
simple kinematic cuts. 
In principle, if one could distinguish between the final state vector boson polarizations, this background 
could be considerably reduced.  A recent paper by Han {\it et al.} argues that this can 
be done \cite{Han:2009em}.  In any case, we find that the cuts we implement reduce this
background to a manageable level.
We therefore define our signal cross section for TeV scale Higgs boson 
production through vector boson scattering as 
\begin{eqnarray}
&&\sigma_{signal}(pp \to V_{L}V_{L}jj) \cr
&&\quad \equiv  \sigma_{total}(pp \rightarrow VVjj) - \sigma_{SM}(pp \rightarrow VVjj),
\label{signal_definition}
\end{eqnarray}
where $\sigma_{SM}(pp \rightarrow VVjj)$ is the cross section for the EW background
process of the Standard Model, and $\sigma_{total}(pp \rightarrow VVjj)$ includes the contributions from the Higgs triplet model under consideration.

The second class of background processes are QCD backgrounds, 
which involve the production of two vector bosons plus additional jets: 
\begin{equation}
pp  \to V V + nj ,
\end{equation}
where $n$ is the number of jets in the final state, typically with $n=2$.  
We also include the 1-jet 
background because we only tag one forward jet in our analysis.  It is possible to tag two high-$p_T$ forward 
jets, but such double 
tagging has been shown to be too costly to the number of signal events \cite{Barger:1990py}.  Therefore, 
we choose to tag a single forward jet in our analysis, as this retains a larger fraction of the signal events.

We evaluate these backgrounds to order $\mathcal{O}(\alpha^2\alpha_s^n)$, where $n$ = 1 or 2, in contrast 
to the irreducible EW background, which is of order $\mathcal{O}(\alpha^4)$.  Since the $VV$ scattering 
signal is a purely electroweak process with no colour exchange and relatively low jet activity in the 
central region, the jets arising in the QCD background are generally more central and have higher $p_{T}$ 
than the spectator jets from the signal events.  Therefore, in addition to a forward jet tag, imposing a 
central jet veto to reject events with hard central jets is very useful in reducing the QCD background.  

The third type of background we consider includes top quarks in the final state, generally along with a vector
boson and possibly an additional jet: 
\begin{eqnarray}
pp & \to & V t \overline{t} , \\
pp & \to & V t \overline{t} j.
\end{eqnarray}
These backgrounds arise from the leptonic decays of the vector boson, $V = W^\pm$ or $Z$, along with
the $t$ and $\bar{t}$ decaying into real $W$ bosons, which then decay leptonically or hadronically.  
The resulting 
final state can have many leptons and jets that can mimic the signal. 
Fortunately, unlike the jets from the signal process, the jets from the $t$-quark background tend to emerge 
in the 
central region of the detector.  Therefore, the $t$-quark background can also be suppressed by imposing a 
forward jet tag and a central jet veto.

In the following subsections we 
give a more detailed description of the backgrounds 
to each of the final states and the kinematic cuts we use
to reduce them to manageable levels.

\subsection{$W^{\pm} W^{\pm}$}

The SM backgrounds to the $W^\pm W^\pm$ channel are:
\begin{itemize}
\item EW Background:
\begin{eqnarray}
pp \to W^\pm W^\pm j j, \:  \mathcal{O}(\alpha^4)
\end{eqnarray}
\item QCD Background:
\begin{equation}
pp \to W^\pm W^\pm j j, \:  \mathcal{O}(\alpha^2 \alpha_s^2)
\end{equation}
\item Top Quark Background:
\begin{eqnarray}
pp &\to& W^\pm t \overline{t} , \\
pp &\to& W^\pm t \overline{t} j,
\end{eqnarray}
with $t \rightarrow W b$.  
\item $W^\pm Z$ Background:
\begin{eqnarray}
pp \rightarrow W^\pm Z + nj  
\end{eqnarray}
with the $W^\pm$ and $Z$ decaying leptonically,  $W^\pm Z\to l^\pm \nu l^+ l^-$.  This process is a background to the $W^\pm W^\pm$ 
signal if  the $l^\mp$ from the Z decay is not observed, due to the finite coverage of the EM calorimeter.  
We therefore include all the EW, QCD and $t$-quark backgrounds to the $W^\pm Z$ channel as backgrounds
to the $W^\pm W^\pm$ channel. 
\end{itemize}

We include the $W^+W^+$ final state as well as 
the charge-conjugate $W^-W^-$ final state to enhance our statistics in this channel.
Note that the cross sections for the two charge-conjugate channels are not identical due to different
parton distribution functions contributing to the different charge states.

For the $W^\pm W^\pm$ channel, the leptonic decay mode in which we are interested is 
$W^{\pm}W^{\pm} \rightarrow l^{\pm} \nu l^{\pm} \nu$, where $l = e, \mu$.  
Because of the neutrinos in the final state, the $W$ boson pair cannot be fully reconstructed.  
Therefore, we consider the cluster transverse mass of the $WW$ pair
\cite{Barger:1983jx,Barger:1987re,Bagger:1995mk} rather than the $WW$
invariant mass distribution.  In general, for a pair of vector bosons decaying leptonically,
we define the cluster transverse mass as
\begin{equation}
M_T^2 (VV) = \left[ E_T^{leptons} +  \not{\! E}_T \right]^2 - \left[ \mathbf{p}_T^{leptons} 
+ \not{\! \mathbf{p}}_T  \right]^2
\label{MT(VV)}
\end{equation}
where $E_T = \sqrt{M^2 + p_T^2}$, and $M$ is the invariant mass.  Therefore, the $WW$ cluster 
transverse mass is given by
\begin{widetext}
\begin{equation}
M_T^2 (WW)  =  
\left[ \sqrt{M^2 (ll) + p_T^2 (ll)}  +
\sqrt{ M^2 (\nu\nu) + \left| \not{\! \mathbf{p}}_T \right|^2} \right]^2 
- \left[ \mathbf{p}_T (ll) +   \not{\! \mathbf{p}}_T  \right] ^ 2 
\end{equation}
\end{widetext}
where $\mathbf{p}_T (ll)=\mathbf{p}_{T}(l_1)+\mathbf{p}_{T}(l_2)$.
However, since the invariant mass of the two neutrinos, $M(\nu\nu)$, 
cannot be reconstructed, we instead define the $WW$ cluster transverse mass as
\begin{eqnarray}
M_T^2 (WW) & = & 
\left[ \sqrt{M^2 (ll) + p_T^2 (ll)} + \left| \not{\! \mathbf{p}}_T \right| \right]^2 
\cr
& & \qquad \qquad \qquad - \left[ \mathbf{p}_T (ll) +   \not{\! \mathbf{p}}_T  \right] ^ 2 .
\label{MT(WW)}
\end{eqnarray}
With this definition, the production of a doubly-charged Higgs boson 
through $W^\pm W^\pm$ scattering results in a broad peak in the $M_T (WW)$ distribution with an endpoint 
at approximately the resonance mass.  This can clearly be seen in Fig.~\ref{fig:w+w+_cuts} for the
$M_\Phi=1$~TeV case, while the broader $\Phi^{\pm\pm}$ width for the $M_\Phi=1.5$~TeV case smears out the 
$M_T (WW)$ distribution.

\subsection{$W^{\pm} Z$}

In the $W^{\pm} Z$ channel, we consider the following SM backgrounds:
\begin{itemize}
\item EW Background:
\begin{equation}
pp \rightarrow W^\pm Z j j, \:  \mathcal{O}(\alpha^4)
\end{equation}
\item QCD Background:
\begin{eqnarray}
pp &\to& W^\pm Z j , \:  \mathcal{O}(\alpha^2 \alpha_s) \\
pp &\to& W^\pm Z j j, \:  \mathcal{O}(\alpha^2 \alpha_s^2)
\end{eqnarray}
\item Top Quark Background:
\begin{eqnarray}
pp &\to& Z t \overline{t} , \\
pp &\to& Z t \overline{t} j,
\end{eqnarray}
with $t \rightarrow W b$.
\end{itemize}
Note that we consider the $W^+Z$ final state as well 
as the charge-conjugate $W^-Z$ final state to enhance our statistics in this channel.

Using Equation ~\ref{MT(VV)}, we obtain the $WZ$ cluster transverse mass: 
\begin{eqnarray}
M_T^2 (WZ) & = & \left[ \sqrt{M^2 (lll) 
+ p_T^2 (lll)} + \left| \not{\! \mathbf{p}}_T \right| \right]^2 \cr
& &  \qquad \qquad \qquad 
- \left[ \mathbf{p}_T (lll) +   \not{\! \mathbf{p}}_T  \right] ^ 2 
\label{MT(WZ)}.
\end{eqnarray}
The $M_T (WZ)$ distribution for singly-charged Higgs boson production 
through $WZ$ scattering shows a distinctive peak at the resonance mass, as seen in Fig.~\ref{fig:wz_cuts}.

\subsection{$W^{+} W^{-}$}

In the $W^{+} W^{-}$ channel, we consider the following SM backgrounds:
\begin{itemize}
\item EW Background:
\begin{equation}
pp \rightarrow W^+ W^- j j, \:  \mathcal{O}(\alpha^4)
\end{equation}
\item QCD Background:
\begin{eqnarray}
pp &\to& W^+ W^- j, \:  \mathcal{O}(\alpha^2 \alpha_s) \\
pp &\to& W^+ W^- j j, \:  \mathcal{O}(\alpha^2 \alpha_s^2)
\end{eqnarray}
\item Top Quark Background:
\begin{eqnarray}
pp &\to& t \overline{t} , \\
pp &\to& t \overline{t} j ,
\end{eqnarray}
with $t \rightarrow W b$.  
\end{itemize}

The $W^\pm Z$ scattering process may also be a background to the $W^+ W^-$ channel 
if the $l^\pm$ from the $Z$ decay is not detected.  However, this is far less important than the 
large QCD and $t$-quark backgrounds to $W^+ W^-$ scattering, so it will be neglected.

As in the $W^\pm W^\pm$ channel, the $W^+W^-$ invariant mass cannot be fully reconstructed 
for leptonic final states, so instead we use the $W^+W^-$ cluster transverse mass, also given by 
Equation \ref{MT(WW)}.

\subsection{$ZZ$}

In the $ZZ$ channel, we consider the following SM backgrounds:
\begin{itemize}
\item EW Background:
\begin{equation}
pp \rightarrow Z Z j j, \:  \mathcal{O}(\alpha^4)
\end{equation}
\item QCD Background:
\begin{eqnarray}
pp &\to& Z Z j , \:  \mathcal{O}(\alpha^2 \alpha_s) \\
pp &\to& Z Z j j, \:  \mathcal{O}(\alpha^2 \alpha_s^2)
\end{eqnarray}
\end{itemize}
An advantage of the $ZZ \rightarrow 4 l$ decay mode is that we can 
completely reconstruct the final state and make use of the $ZZ$ invariant mass distribution in our results.

\subsection{Kinematic Cuts}
The kinematic cuts used to reduce the SM backgrounds to vector boson scattering are well 
established \cite{Barger:1990py,Bagger:1993zf,Bagger:1995mk,Kilgore:1996pj}.  Our values 
for the cuts are summarized in Table~\ref{tab:cuts}.  We impose cuts on the $p_T$ of all 
charged leptons, the missing transverse momentum $({\not p}_T)$ in the $W^{\pm}Z$ channel,  
and impose back-to-back lepton cuts ($\Delta y(ll)$ and $\Delta p_T(ll)$) in the $W^{\pm}W^{\pm}$ and $W^+W^-$ channels.  
We impose rapidity cuts on the final state leptons of $|y(l)| < 2.5$ and on 
the tag jet of $3.0 < |y(j_{tag})| < 5.0$ to take into account detector acceptances.  
Imposing a central jet veto was not necessary in the $ZZ$ channel, as this would further 
reduce the already small number of signal events.  Note that the central jet veto cuts are 
to be interpreted as: ``Reject all events with jets having $|y| < y_{max}$ and $p_{T} > {p_T}_{min}$".  
It should also be noted that, although we tag a single energetic forward jet with high-$p_T$, all 
jets are assumed to have $p_T > 10$ GeV by default.  This is required in order for MadGraph to 
achieve stable results in its calculations.  We find that these choices of cuts are effective 
in improving the signal to background ratio.

\begin{table}[t]
\begin{center}
\caption{Leptonic cuts and jet cuts used in vector boson scattering to enhance the signal to background ratio. 
 Note that the central jet veto cuts are to be interpreted as: ``Reject all events with jets having 
$|y| < y_{max}$ and $p_{T} > {p_T}_{min}$."}
\begin{ruledtabular}
\begin{tabular}{ll}
Leptonic Cuts & Jet Cuts\\
\hline
\multicolumn{2}{c}{$W^\pm W^\pm$} \\
\hline
$|y(l)| < 2.5$ & $3.0 < |y(j_{tag})| < 5.0$ \\
$p_{T}(l_1) > 200 \; \rm GeV$ & $p_{T}(j_{tag}) > 40 \; \rm GeV$ \\
$p_{T}(l_2) >  50 \; \rm GeV$  & $E(j_{tag}) > 500 \; \rm GeV$ \\
$\Delta p_{T}(ll) =|p_{T}(l_1)-p_{T}(l_2)| > 300 \; \rm GeV $ & $|y(j_{veto})| < 3.0$ \\
$\Delta y(ll) =|y(l_1)-y(l_2)|<3$  & $p_{T}(j_{veto}) > 100 \; \rm GeV$ \\
$M_{T}(WW) > 550 \; (800) \; \rm GeV$ \\
\quad \quad for $M_{\Phi} = 1.0 \; (1.5) \; \rm TeV$  \\

\hline
\multicolumn{2}{c}{$W^\pm Z$} \\
\hline
$|y(l)| < 2.5$ & $3.0 < |y(j_{tag})| < 5.0$ \\
$p_{T}(l_1) > 150 \; \rm GeV$ & $p_{T}(j_{tag}) > 40 \; \rm GeV$ \\
$p_{T}(l_2) > 50 \; \rm GeV$ & $E(j_{tag}) > 500 \; \rm GeV$ \\
$p_{T}(l_3) > 50 \; \rm GeV$  & $|y(j_{veto})| < 3.0$ \\
$\not{p}_{T} > 50 \; \rm GeV$ & $p_{T}(j_{veto}) > 100 \; \rm GeV$ \\
$M_{T}(WZ) > 900 \; (1250) \; \rm GeV$ \\
\quad \quad for $M_{\Phi} = 1.0 \; (1.5) \; \rm TeV$  \\

\hline
\multicolumn{2}{c}{$W^+W^-$} \\
\hline
$|y(l)| < 2.5$ & $3.0 < |y(j_{tag})| < 5.0$ \\
$p_{T}(l_1) > 300 \; \rm GeV$ & $p_{T}(j_{tag}) > 40 \; \rm GeV$ \\
$p_{T}(l_2) >  150 \; \rm GeV$  & $E(j_{tag}) > 800 \; \rm GeV$ \\
$\Delta p_{T}(ll) =|p_{T}(l_1)-p_{T}(l_2)| > 500 \; \rm GeV $ & $|y(j_{veto})| < 3.0$ \\
$\Delta y(ll) =|y(l_1)-y(l_2)|<3$  & $p_{T}(j_{veto}) > 60 \; \rm GeV$ \\
$M_{T}(WW) > 700 \; (1000) \; \rm GeV$ \\
\quad \quad for $M_{\Phi} = 1.0 \; (1.5) \; \rm TeV$  \\

\hline
\multicolumn{2}{c}{$ZZ$} \\
\hline
$|y(l)| < 2.5$ & $3.0 < |y(j_{tag})| < 5.0$ \\
$p_{T}(l) > 70 \; \rm GeV$ & $p_{T}(j_{tag}) > 40 \; \rm GeV$ \\
 $M(ZZ) > 900 \; (1250) \; \rm GeV$  & $E(j_{tag}) > 500 \; \rm GeV$ \\
\quad \quad for $M_{\Phi} = 1.0 \; (1.5) \; \rm TeV$  & no jet veto\\

\end{tabular}
\end{ruledtabular}
\label{tab:cuts}
\end{center}
\end{table}

\section{Results}

We used the MadGraph software package \cite{Alwall:2007st}, version 4.2.6, 
to generate weighted signal and background events.
To obtain our results, we used the following SM input parameters \cite{Amsler:2008zzb}: 
$G_{F} = 1.16637 \times 10^{-5} \; \rm GeV^{-2}$, $M_{Z} = 91.188$ GeV, $\alpha(M_{Z}) = 1/127.9$, 
$\alpha_{s}(M_{Z}) = 0.118$, $M_{t} = 171.2$ GeV.  We assumed a SM Higgs boson mass of
$M_{h} = 120$ GeV and used the CTEQ6M parton distribution functions \cite{Pumplin:2002vw}. 
We assumed the nominal LHC energy of $\sqrt{s} = 14$ TeV and an integrated 
luminosity of $\mathcal{L} = 100 \;\mathrm{fb}^{-1}$ to obtain our results. 

In the Higgs triplet models that we studied, there are additional parameters in the scalar sector that 
must be defined.  For all models, we use triplet masses of $M_{\Phi} = 1.0$ and 1.5~TeV and a triplet vev of $v^\prime = 39$ GeV; as discussed in Section II, this value of the triplet 
vev corresponds to the upper bound in the Georgi-Machacek model.  In the Littlest Higgs model, the symmetry breaking scale, $f$, is an additional parameter that we set to $f = 2$ TeV.  With these parameter values, we used the 
BRIDGE \cite{Meade:2007} package of MadGraph to calculate the scalar triplet decay widths for each of the models, and the resulting values are given in Table \ref{tab:widths}.

   \begin{figure*}[t]
   \begin{center}
   \includegraphics[width=3.5in]{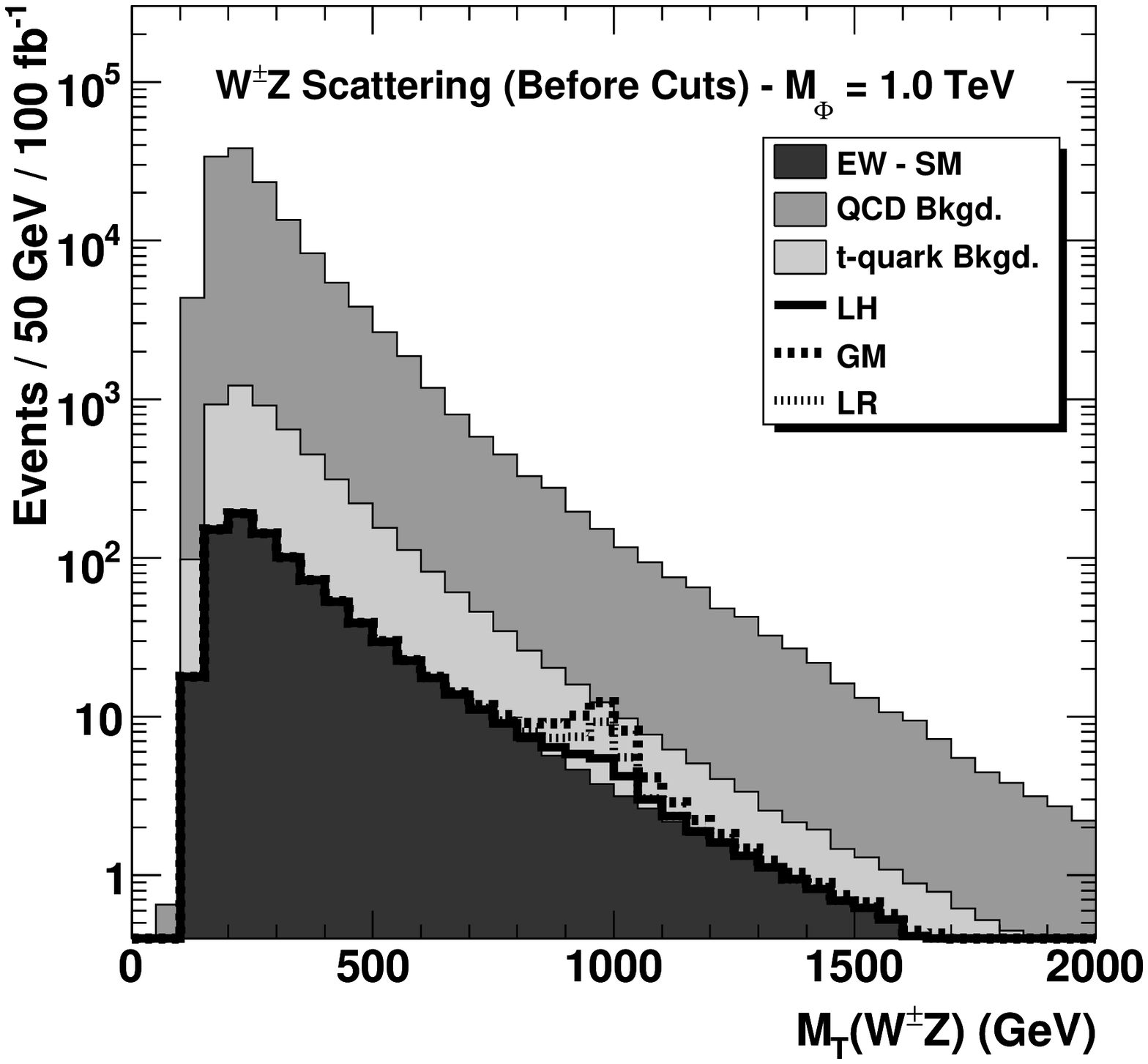}
   \includegraphics[width=3.5in]{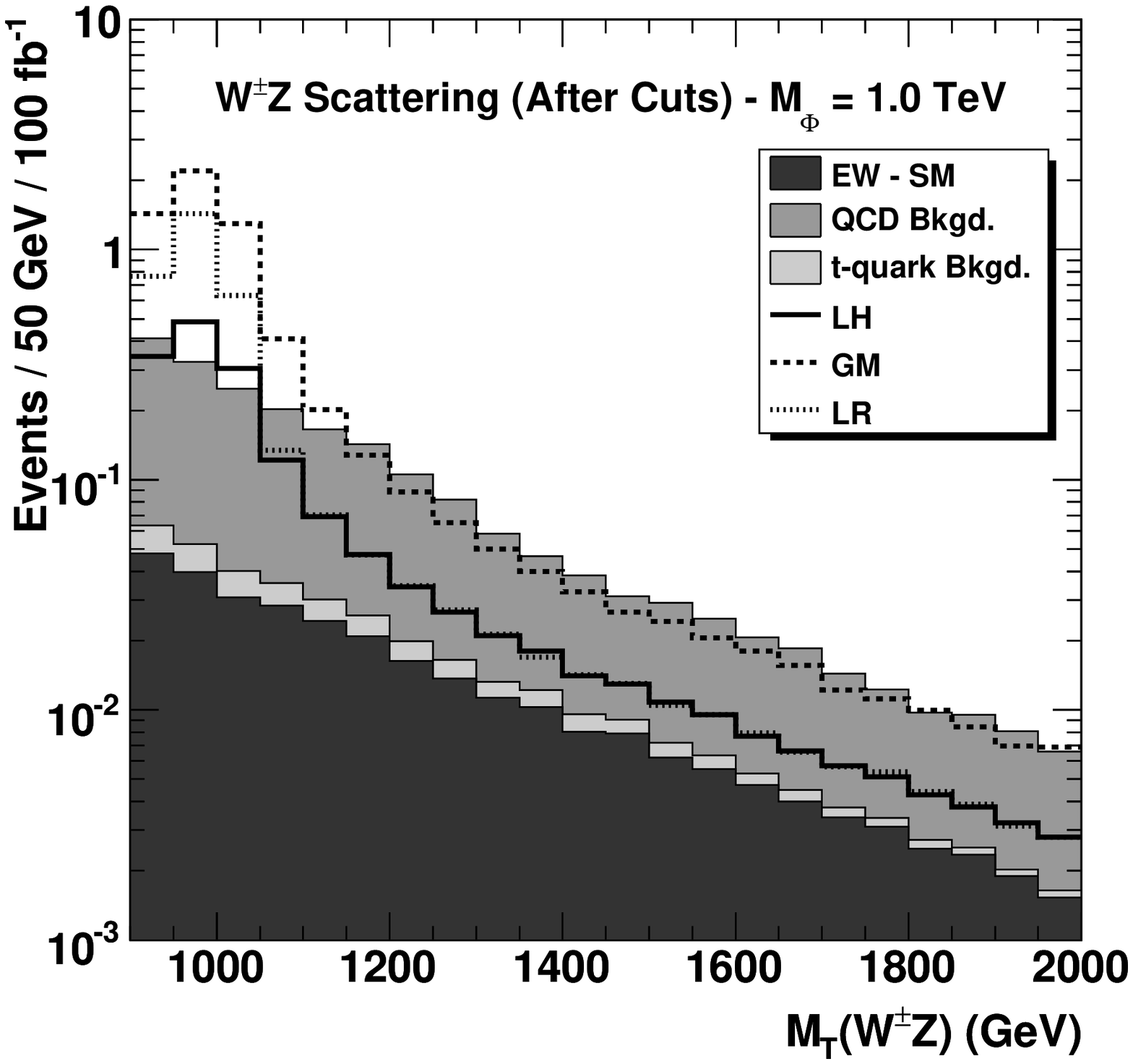}
   \caption{The transverse mass distributions for the $W^{\pm}Z$ signals 
of the Littlest Higgs (LH), Georgi-Machacek (GM) and Left-Right Symmetric (LR) models, 
along with the backgrounds, before and after imposing the cuts of Table~\ref{tab:cuts}.  
A Higgs triplet mass of $M_{\Phi} = 1.0$ TeV was used, along with a triplet vev of $v' = 39$ GeV, 
assuming an integrated luminosity of $\mathcal{L} = 100 \;\mathrm{fb}^{-1}$.  Note that the backgrounds are stacked, whereas the signal lines are not.  
}
   \label{fig:backgrounds}
   \end{center}
   \end{figure*}  

For the purpose of discussion we start with $W^\pm Z$ scattering as a representative case.  The
process is  $pp \rightarrow W^{\pm}Z jj$ with, for the ``gold-plated'' leptonic decay modes,
$W^{\pm} \rightarrow l^{\pm} \nu$ and $Z \rightarrow l^{+} l^{-} \; (l = e, \mu)$.
The experimental signature is given by two high-$p_T$ forward jets, along with three high-$p_T$ charged leptons in the central 
region of the detector.  This is accompanied by a large amount 
of missing $p_T$ attributed to the undetected neutrino.  
The transverse mass distributions for each type of 
background along with the  $W^\pm Z$ signal before and after imposing the cuts of Table~\ref{tab:cuts} are shown in Fig.~\ref{fig:backgrounds}, for a triplet mass of $M_{\Phi} = 1.0$ TeV.
It can be seen from this figure that our choice of cuts reduces the irreducible EW
background to a manageable level and substantially reduces the QCD and $t$-quark backgrounds.

   \begin{figure*}[tp]
  \begin{center}
 \includegraphics[width=3.5in]{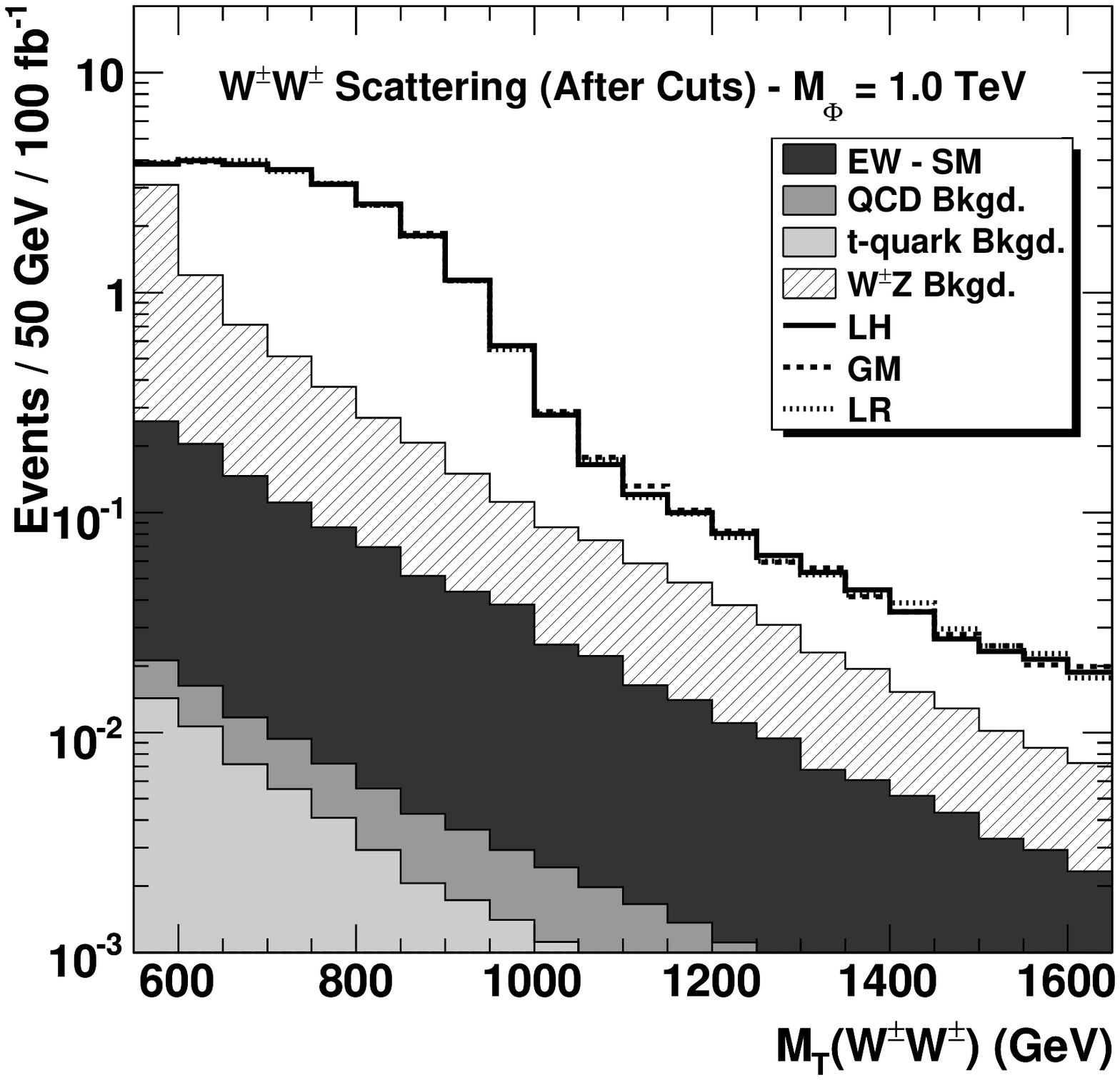}
 \includegraphics[width=3.5in]{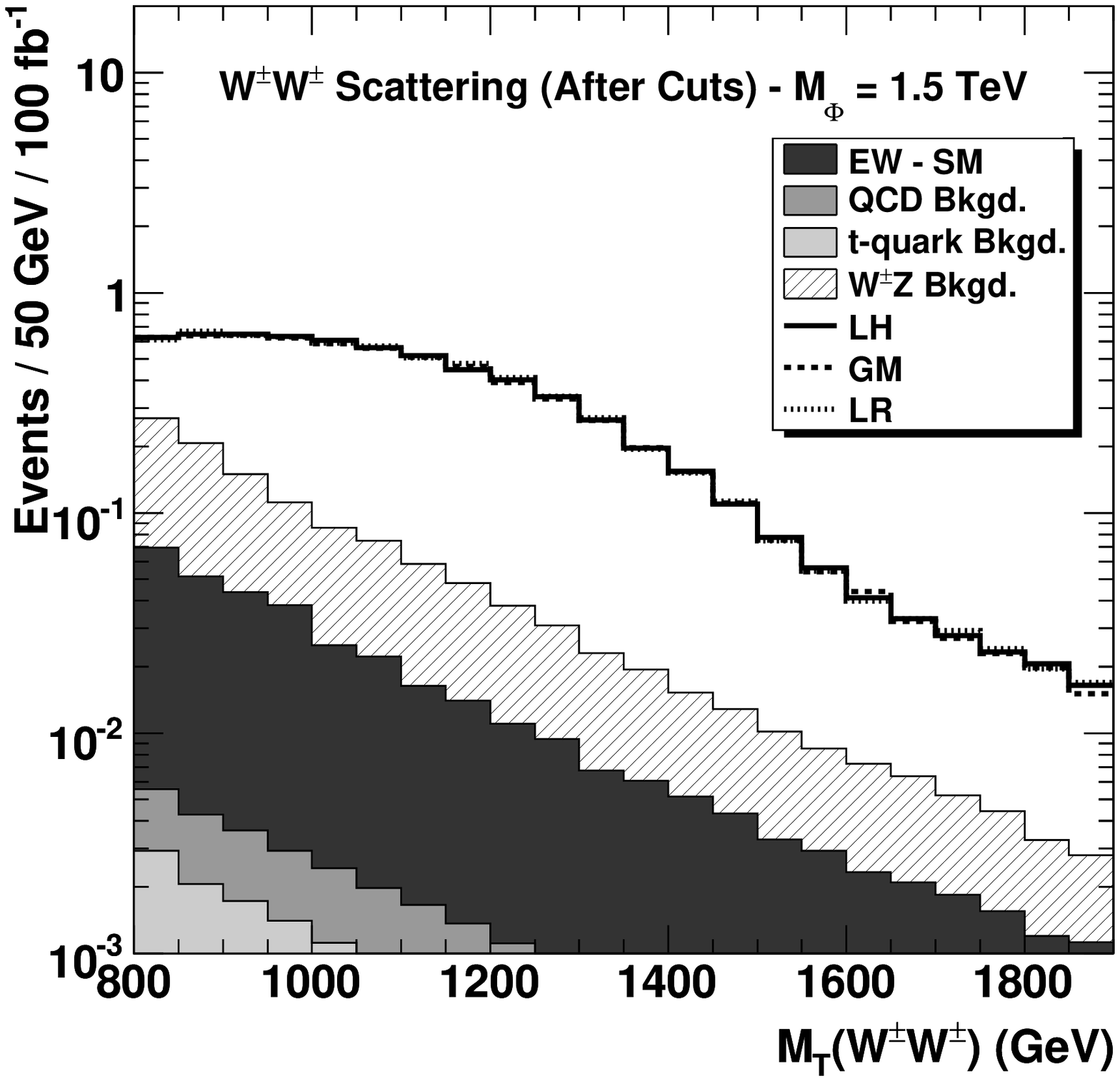}
 \caption{The transverse mass distributions for $W^{\pm}W^{\pm}$ scattering in the 
Littlest Higgs (LH), Georgi-Machacek (GM) and Left-Right Symmetric (LR) models, along with 
the backgrounds, after imposing the cuts of Table~\ref{tab:cuts}.  Higgs triplet 
masses of $M_{\Phi} = 1.0$ and 1.5  TeV were used, along with a triplet vev of $v' = 39$ GeV, 
assuming an integrated luminosity of $\mathcal{L} = 100 \;\mathrm{fb}^{-1}$.  Note that the backgrounds are stacked, whereas the signal lines are not.}
\label{fig:w+w+_cuts}
\end{center}
\end{figure*}      

   \begin{figure*}[bp]
  \begin{center}
 \includegraphics[width=3.5in]{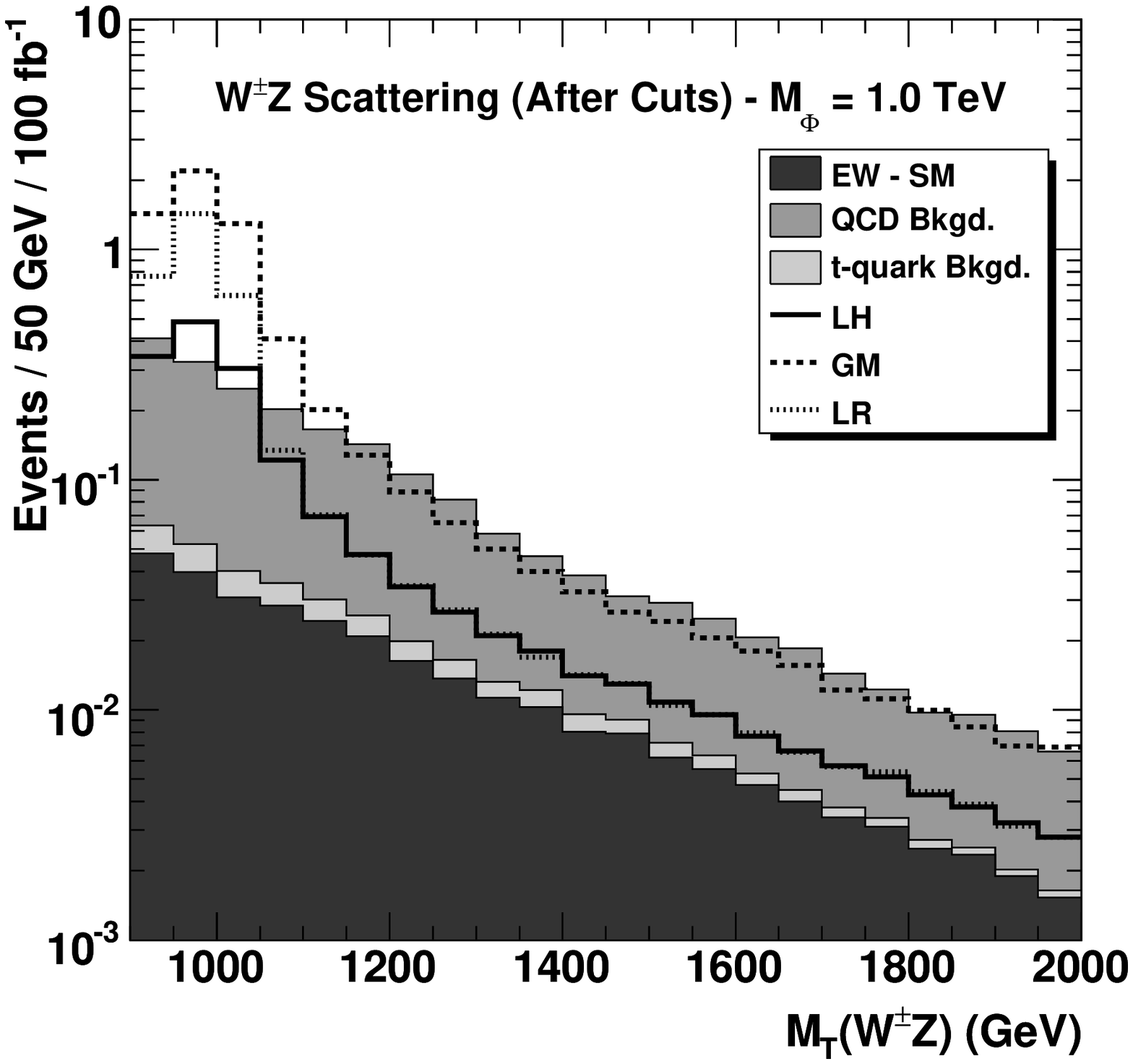}
 \includegraphics[width=3.5in]{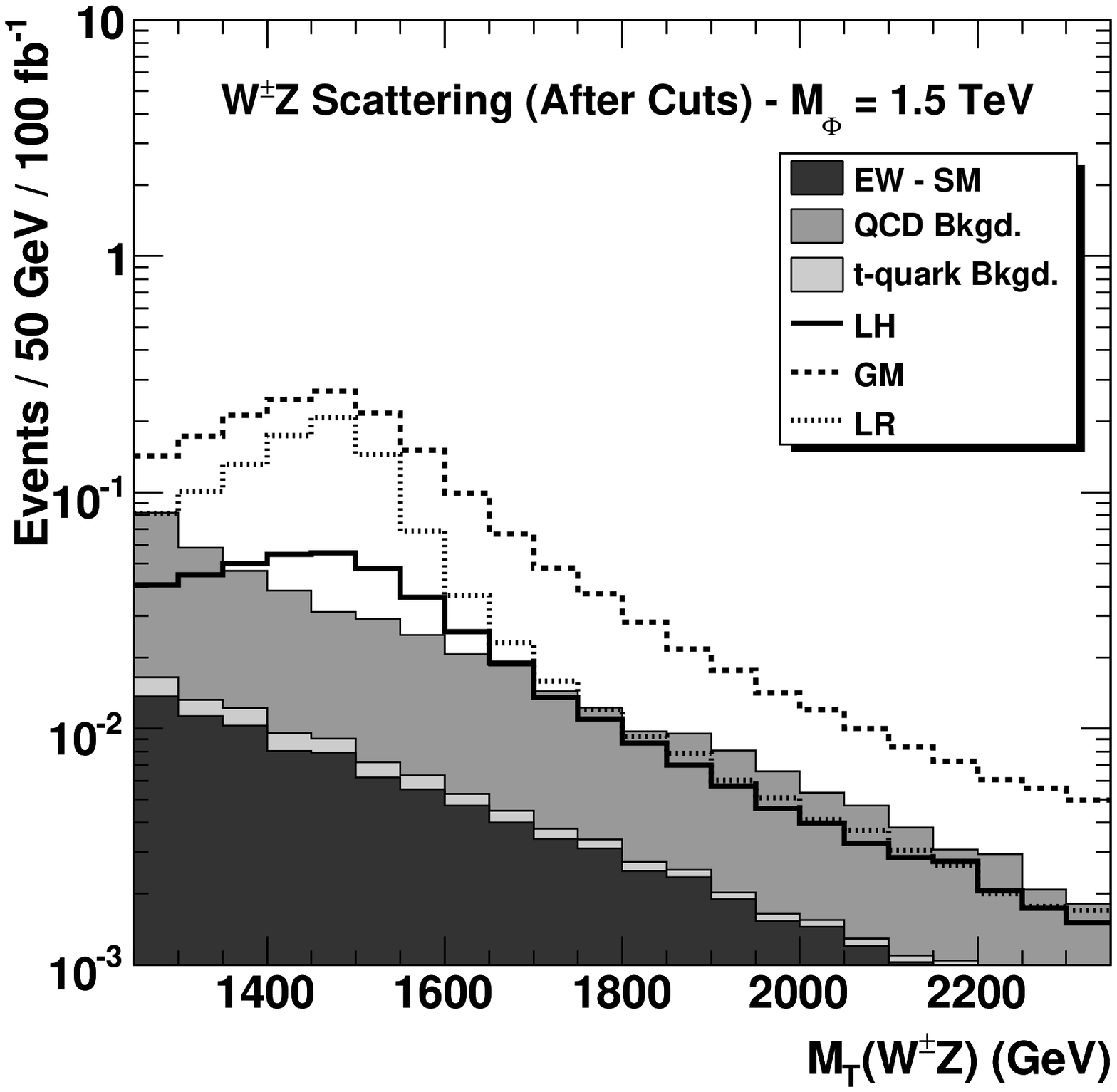}
 \caption{The transverse mass distributions for $W^{\pm}Z$ scattering in 
the Littlest Higgs (LH), Georgi-Machacek (GM) and Left-Right Symmetric (LR) models, 
along with the backgrounds, after imposing the cuts of Table~\ref{tab:cuts}.  
Higgs triplet masses of $M_{\Phi} = 1.0$ and 1.5  TeV were used, along with a triplet 
vev of $v' = 39$ GeV, assuming an integrated luminosity of $\mathcal{L} = 100 \;\mathrm{fb}^{-1}$.  Note that the backgrounds are stacked, whereas the signal lines are not.}
\label{fig:wz_cuts}
\end{center}
\end{figure*}    

   \begin{figure*}[tp]
  \begin{center}
 \includegraphics[width=3.5in]{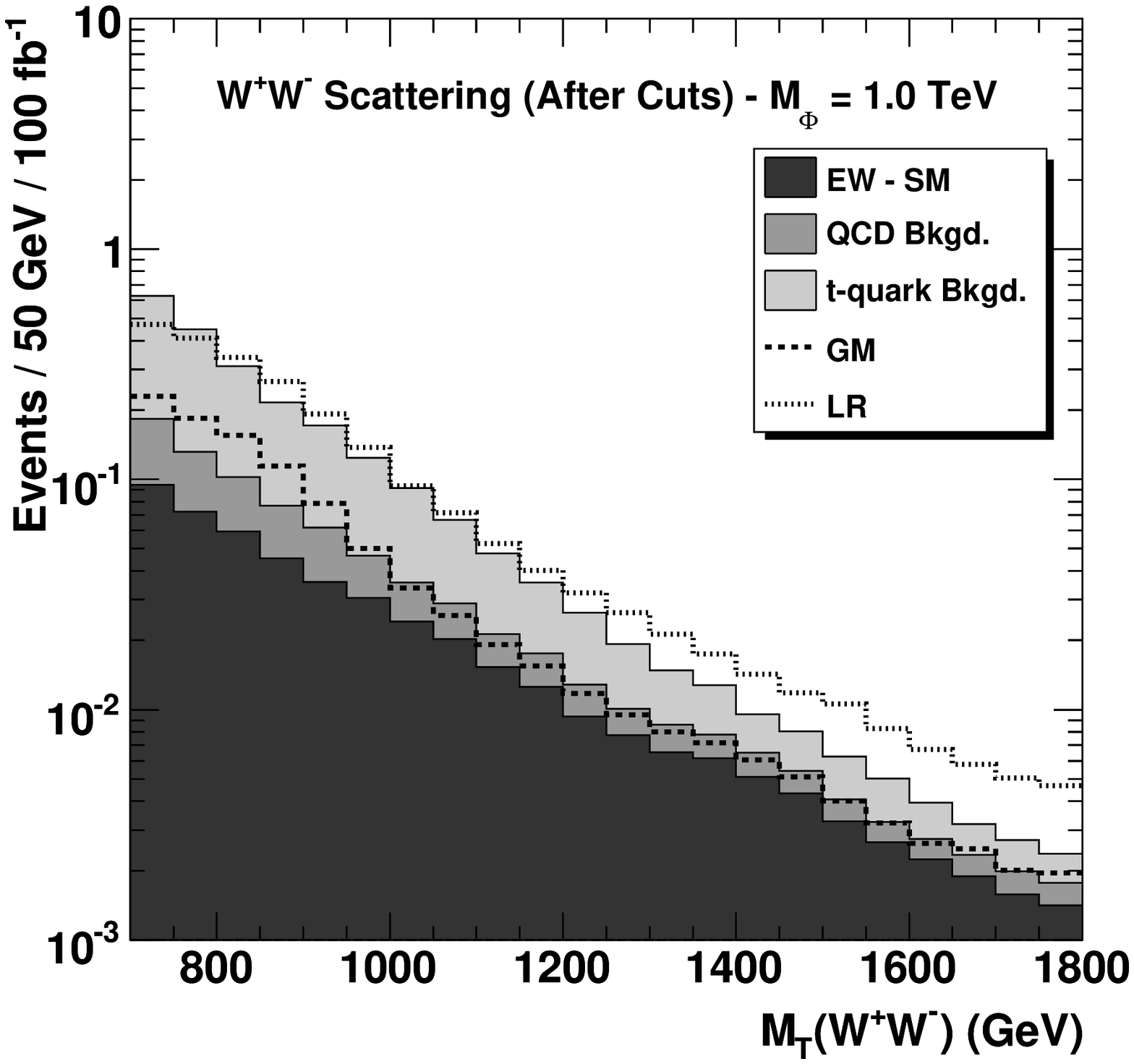}
 \includegraphics[width=3.5in]{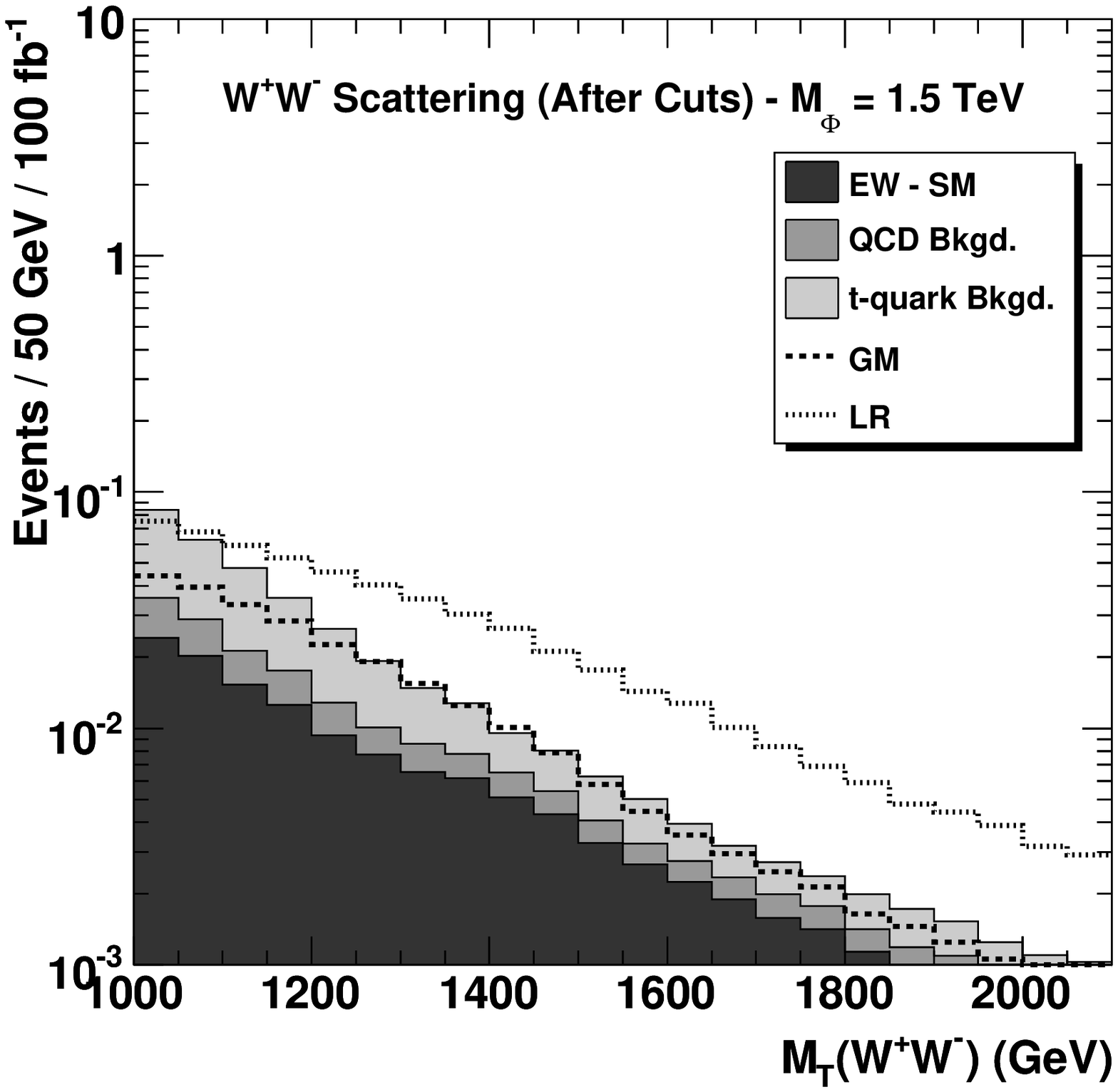}
 \caption{The transverse mass distributions for $W^{+}W^{-}$ scattering in the Georgi-Machacek 
(GM) and Left-Right Symmetric (LR) models (this signal does not exist in the Littlest Higgs model
at lowest order), along with the backgrounds, after imposing the cuts of 
Table~\ref{tab:cuts}.  Higgs triplet masses of $M_{\Phi} = 1.0$ and 1.5  TeV were used, along with a triplet vev of $v' = 39$ GeV, 
assuming an integrated luminosity of $\mathcal{L} = 100 \;\mathrm{fb}^{-1}$.  Note that the backgrounds are stacked, whereas the signal lines are not.}
\label{fig:w+w-_cuts}
\end{center}
\end{figure*}    

   \begin{figure*}[bp]
  \begin{center}
 \includegraphics[width=3.5in]{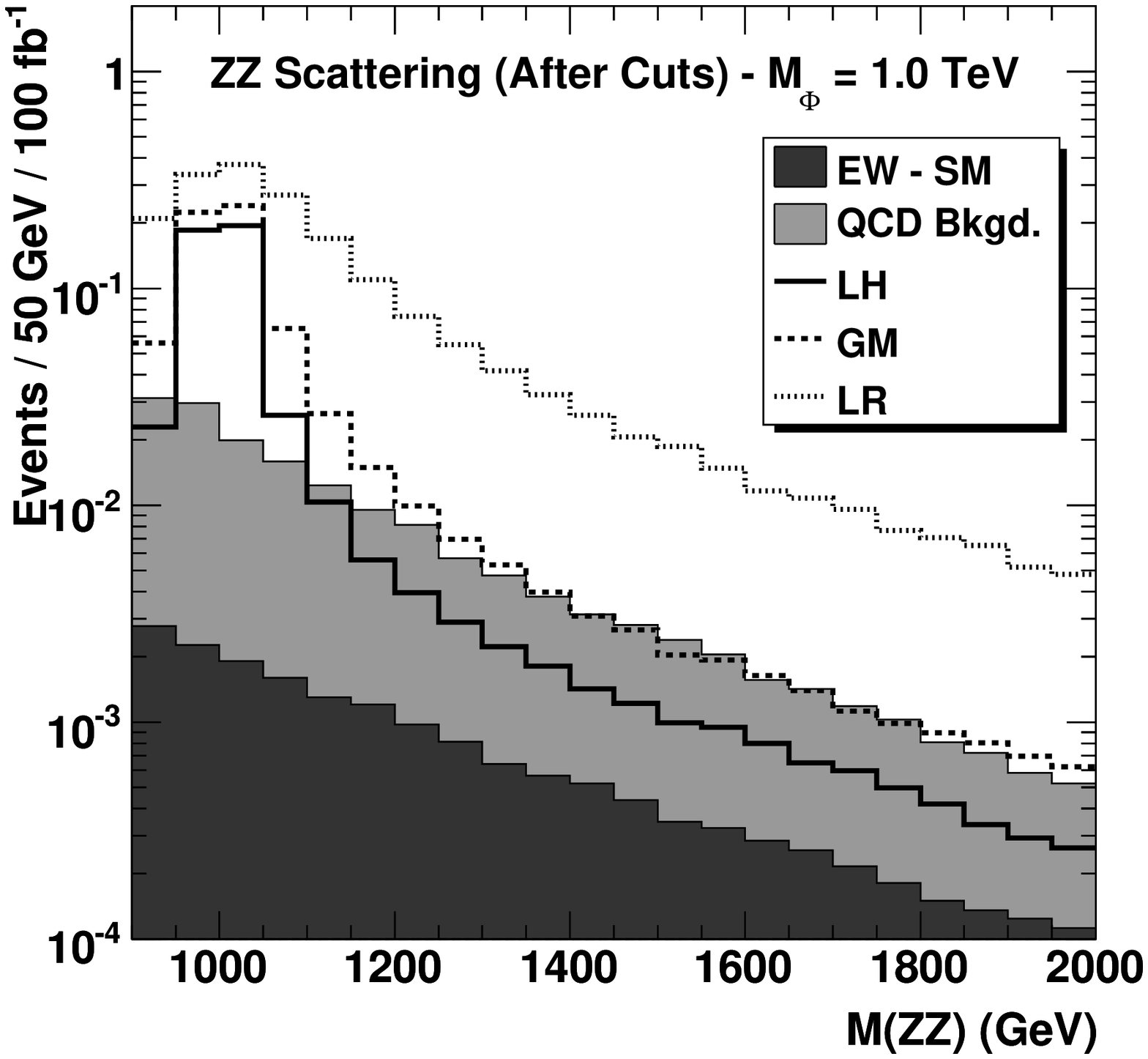}
 \includegraphics[width=3.5in]{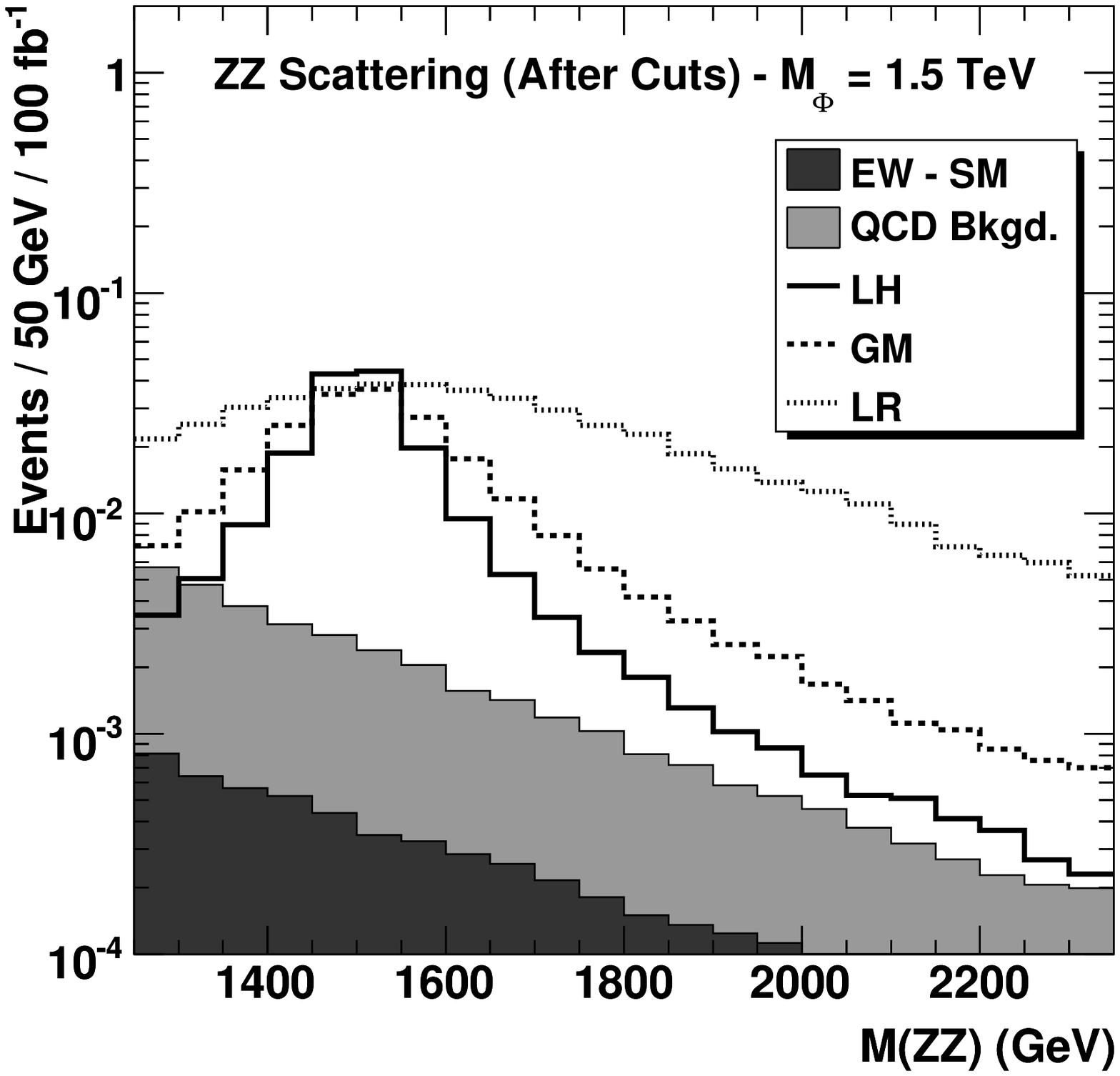}
 \caption{The invariant mass distributions for $ZZ$ scattering 
in the Littlest Higgs (LH), Georgi-Machacek (GM) and Left-Right Symmetric (LR) models, 
along with the backgrounds, after imposing the cuts of Table~\ref{tab:cuts}.  
Higgs triplet masses of $M_{\Phi} = 1.0$ and 1.5  TeV were used, along with a triplet 
vev of $v' = 39$ GeV, assuming an integrated luminosity of $\mathcal{L} = 100 \;\mathrm{fb}^{-1}$.  Note that the backgrounds are stacked, whereas the signal lines are not.}
\label{fig:zz_cuts}
\end{center}
\end{figure*}    

All channels show qualitatively similar improvements in the signal to background ratio, so
for the remaining channels we simply show the resulting mass distributions
after imposing the kinematic cuts given in Table~\ref{tab:cuts}.
The resulting cluster transverse mass distributions 
for the $W^\pm W^\pm$, $W^\pm Z$, and $W^+W^-$ channels, and the invariant mass distribution 
for the $ZZ$ channel are shown 
in Figs.~\ref{fig:w+w+_cuts}-\ref{fig:zz_cuts} for $M_\Phi = 1.0$ and 1.5~TeV.  

To better understand these plots, it is helpful to refer to the scalar triplet partial widths in these models,
which are given in Table~\ref{tab:widths}.  
We start with the $W^\pm W^\pm$ channel shown in Fig.~\ref{fig:w+w+_cuts}.
The transverse mass distributions for all three models in the  $W^\pm W^\pm$ channel
lie on top of each other because the doubly-charged member of the scalar triplet has identical
$W^+W^+\Phi^{--}$ vertices and the same decay width in each model.
In contrast, the $W^{\pm}Z$ channel,  shown in Fig.~\ref{fig:wz_cuts}, exhibits differences between all three models.
The peak width for the Littlest Higgs model signal is larger than that of the Left-Right Symmetric 
model signal because, although they have the same $W^+Z \Phi^-$ vertex, the $\Phi^\pm$ of the Littlest Higgs model can also 
decay to $ W^\pm h$, resulting in a larger $\Phi^{\pm}$ width.  Furthermore, although the $H_5^\pm$ of the Georgi-Machacek model can only decay to $W^\pm Z$, as discussed previously, the $W^+ ZH_5^-$ coupling is larger by a 
factor of $\sqrt{2}$ compared to the $W^+ Z\Delta_L^-$ coupling in the Left-Right Symmetric model, resulting
in a $H_5^\pm$ width that is a factor of two larger than the $\Delta_L^\pm$ width and also a larger production
cross section. 
Finally, the differential cross sections in the $ZZ$ channel shown in Figure \ref{fig:zz_cuts}
differ quite significantly due to the 
different $\Phi^0 ZZ$ couplings and $\Phi^0$ widths.  The width  of the neutral member of the scalar
triplet is smallest in the Littlest Higgs model, partially 
due to the absence of the $\Phi^0 \to W^+W^-$ decay mode at leading order in this model, but also due to the
value of the $\Phi^0 ZZ$ coupling.  In both the GM and LR models the $\Phi^0$ can also decay $W^+W^-$ which
increases the total width.  These effects are not so apparent in the $W^+W^-$ channel because the signal
is more deeply buried in the background.

\begin{table*}[t]
\begin{center}
\caption{The partial widths of the Higgs triplet bosons for the models studied in this paper. The widths 
are in GeV.
\label{tab:widths}}
\begin{ruledtabular}
\begin{tabular}{lcccccc}
	 & \multicolumn{2}{c}{Littlest Higgs} & \multicolumn{2}{c}{Georgi-Machacek} & \multicolumn{2}{c}{Left-Right Symmetric} \\
	& $M_\Phi=1.0$ TeV & $M_\Phi=1.5$ TeV & $M_\Phi=1.0$ TeV & $M_\Phi=1.5$ TeV & 
	$M_\Phi=1.0$ TeV & $M_\Phi=1.5$ TeV \\
Decay 	& (GeV) & (GeV) & (GeV) & (GeV) & (GeV) & (GeV) \\
\hline
$\Gamma(\Phi^{\pm\pm}\to W^\pm W^\pm)$
			& 64	& 220	& 64	& 220	& 64	& 220 \\
$\Gamma(\Phi^\pm\to W^\pm Z)$
			& 32	& 110	& 63 & 220	& 32	& 110 \\
$\Gamma(\Phi^\pm\to W^\pm h)$
			& 42	& 150	& -	& -		& -	& -	\\
$\Gamma(\Phi^0\to W^+W^-)$
			&  - 	&  -		& 21	&  73		& 64	& 220  \\
$\Gamma(\Phi^0\to ZZ)$ 	
			& 31	& 110	& 42	& 150	& 126 & 440	\\  
\end{tabular}
\end{ruledtabular}
\end{center}
\end{table*}

More quantitatively, in Table~\ref{tab:xsect} we give the cross sections for the signal and 
backgrounds after imposing the cuts of Table~\ref{tab:cuts}.  
We show cross sections for the Littlest Higgs model, the Georgi-Machacek model, and the
Left-Right Symmetric model for scalar triplet masses of $M_\Phi = 1.0$ and 1.5~TeV.  The row labelled as 
EW-SM is
the Standard Model cross section, assuming a Higgs mass of $M_h = 120$ GeV, which we refer to in the text as
the irreducible EW background.  The rows labelled as Signal-LH,
Signal-GM, and Signal-LR are the difference between the EW-SM cross section and the EW-LH, EW-GM,
and EW-LR cross sections respectively, as defined by Equation \ref{signal_definition}.

\begin{table*}[bp]
\begin{center}
\caption{LHC cross sections (in fb) for vector boson scattering using the cuts listed in Table 
\ref{tab:cuts}. The cross sections are shown for the three models we are
considering, using a triplet vev of $v^\prime = 39$ GeV and Higgs triplet masses of $M_\Phi = 1.0$ TeV (left-hand columns) 
and $M_\Phi = 1.5$ TeV  (right-hand columns).  The cross section values obtained using MadGraph are accurate to $\sim1\%$.}
\begin{ruledtabular} 
\begin{tabular}{llllclll}
 &  \multicolumn{3}{c}{$M_\Phi=1.0$ TeV}  & \vline &   \multicolumn{3}{c}{$M_\Phi=1.5$ TeV}  \\
\hline
Process 			&  Leptonic Cuts 	& + Jet Tag 	& + Jet Veto   
		& \vline 	& Leptonic Cuts 	& + Jet Tag	& + Jet Veto \\
\hline
$\bf W^\pm W^\pm$ 	& 					& 					& 		
		& \vline 	&					& 					& 		\\
\hline
EW - LH 			& $9.39\times 10^{-1} $ 	& $3.03 \times 10^{-1}$ 	& $2.55\times 10^{-1}$ 
		 & \vline  	& $2.92\times 10^{-1}$	& $1.05\times 10^{-1}$	& $8.85\times 10^{-2}$	\\
EW - GM  			& $9.41\times 10^{-1}$	& $3.06\times 10^{-1}$	& $2.56\times 10^{-1}$	
		& \vline 	& $2.88\times 10^{-1}$	& $1.03\times 10^{-1}$	& $8.75\times 10^{-2}$	\\
EW - LR   			& $9.36\times 10^{-1}$ 	& $3.04\times 10^{-1}$	& $2.56\times 10^{-1}$
		& \vline 	& $2.91\times 10^{-1}$	& $1.04\times 10^{-1}$	& $8.87\times 10^{-2}$	\\
EW - SM    		& $2.01\times 10^{-1}$	& $1.57\times 10^{-2}$	& $1.06\times 10^{-2}$
		& \vline 	& $6.48\times 10^{-2}$ 	& $4.91\times 10^{-3}$  	& $3.42\times 10^{-3}$	\\
QCD Bkgd.  		& $1.16\times 10^{-1}$	& $7.90\times 10^{-4}$	& $4.22\times 10^{-4}$
		& \vline  	& $4.51\times 10^{-2}$	& $2.47\times 10^{-4}$	& $1.67\times 10^{-4}$ 	\\
$t$-quark Bkgd. 	& $4.42\times 10^{-1}$	& $8.15\times 10^{-4}$	& $5.52\times 10^{-4}$
		& \vline 	& $1.29\times 10^{-1}$	& $1.38\times 10^{-4}$	& $7.33\times 10^{-5}$	\\
$W^{\pm}Z$ Bkgd.  	& $5.75$				& $8.58\times 10^{-2}$	& $6.82\times 10^{-2}$
		& \vline 	& 1.45				& $1.91\times 10^{-2}$ 	& $1.67\times 10^{-2}$ 	\\
Total Bkgd. 		& 6.51				& $1.03\times 10^{-1}$	& $7.98\times 10^{-2}$
		& \vline 	& 1.69				& $2.44\times 10^{-2}$	& $2.04\times 10^{-2}$	\\
Signal - LH  		& $7.38\times 10^{-1}$	& $2.87\times 10^{-1}$	& $2.44\times 10^{-1}$
		& \vline 	& $2.27\times 10^{-1}$ 	& $1.00\times 10^{-1}$	& $8.51\times 10^{-2}$ \\
Signal - GM  		& $7.40\times 10^{-1}$	& $2.90\times 10^{-1}$	& $2.45\times 10^{-1}$
		& \vline 	& $2.23\times 10^{-1}$	& $0.98\times 10^{-1}$	& $8.41\times 10^{-2}$ 	\\
Signal - LR   		& $7.35\times 10^{-1}$	& $2.88\times 10^{-1}$	& $2.45\times 10^{-1}$
		& \vline 	& $2.26\times 10^{-1}$	& $0.99\times 10^{-1}$	& $8.53\times 10^{-2}$	\\

\hline
$\bf W^\pm Z$ 		& 					& 					& 			
	 & 	\vline 	& 					& 					& 		\\
\hline
EW - LH  			& $9.50\times 10^{-2}$ 	& $1.95\times 10^{-2}$	& $1.56\times 10^{-2}$
	&	\vline  	& $3.12\times 10^{-2}$	& $6.84\times 10^{-3}$	& $5.48\times 10^{-3}$	\\
EW - GM 			& $2.44\times 10^{-1}$	& $7.46\times 10^{-2}$ 	& $6.15\times 10^{-2}$
	&	\vline 	& $7.86\times 10^{-2}$	& $2.66\times 10^{-2}$	& $2.22\times 10^{-2}$ \\
EW - LR   			& $1.51\times 10^{-1}$ 	& $4.01\times 10^{-2}$	& $3.28\times 10^{-2}$
	&	\vline  	& $5.13\times 10^{-2}$	& $1.51\times 10^{-2}$	& $1.25\times 10^{-2}$	\\
EW - SM     		& $5.47\times 10^{-2}$ 	& $4.47\times 10^{-3}$	& $3.03\times 10^{-3}$
	&	\vline 	& $1.87\times 10^{-2}$	& $1.60\times 10^{-3}$	& $1.06\times 10^{-3}$	\\
QCD Bkgd.  		& 1.60 				& $1.96\times 10^{-2}$	& $1.72\times 10^{-2}$
	&	\vline  	& $3.34\times 10^{-1}$	& $2.60\times 10^{-3}$ 	& $2.25\times 10^{-3}$	\\
$t$-quark Bkgd.   	& $2.54\times 10^{-1}$ 	& $1.02\times 10^{-3}$	& $7.30\times 10^{-4}$
	&	\vline  	& $5.62\times 10^{-2}$	& $1.07\times 10^{-4}$	& $8.93\times 10^{-5}$	\\
Total Bkgd. 		& 1.91 				& $2.51\times 10^{-2}$	& $2.10\times 10^{-2}$
	&	\vline  	& $4.09\times 10^{-1}$	& $4.31\times 10^{-3}$	& $3.40\times 10^{-3}$	\\
Signal - LH  		& $4.03\times 10^{-2}$ 	& $1.50\times 10^{-2}$	& $1.26\times 10^{-2}$
 	&	\vline 	& $1.25\times 10^{-2}$	& $5.24\times 10^{-3}$	& $4.42\times 10^{-3}$	\\
Signal - GM 		& $1.89\times 10^{-1}$ 	& $7.01\times 10^{-2}$ 	& $5.85\times 10^{-2}$
	&	\vline  	& $5.99\times 10^{-2}$ 	& $2.50\times 10^{-2}$	& $2.11\times 10^{-2}$	\\
Signal - LR 		& $0.96\times 10^{-1}$ 	& $3.56\times 10^{-2}$	& $2.98\times 10^{-2}$
	&	\vline  	& $3.26\times 10^{-2}$ 	& $1.35\times 10^{-2}$	& $1.14\times 10^{-2}$	\\

\hline
$\bf W^+W^-$ 		& 					& 					& 
	&	\vline 	& 					& 					& 		\\
\hline
EW - LH 			& - 					& -					& -		
	&	\vline 	& - 					& -					& -		\\
EW - GM   		& $1.62\times 10^{-1}$ 	& $2.08\times 10^{-2}$	& $9.91\times 10^{-3}$
	&	\vline 	& $7.28\times 10^{-2}$	& $9.20\times 10^{-3}$	& $4.35\times 10^{-3}$	\\
EW - LR   			& $2.35\times 10^{-1}$ 	& $4.63\times 10^{-2}$	& $2.28\times 10^{-2}$
	&	\vline 	& $9.70\times 10^{-2}$	& $1.84\times 10^{-2}$	& $9.23\times 10^{-3}$	\\
EW - SM    		& $1.32\times 10^{-1}$ 	& $1.07\times 10^{-2}$ 	& $4.75\times 10^{-3}$
	&	\vline 	& $5.98\times 10^{-2}$	& $4.32\times 10^{-3}$	& $1.79\times 10^{-3}$	\\
QCD Bkgd.  		& 4.72 				& $1.61\times 10^{-2}$	& $4.29\times 10^{-3}$
	 &	\vline 	& $9.18\times 10^{-1}$	& $3.13\times 10^{-3}$	& $8.34\times 10^{-4}$	\\
$t$-quark Bkgd.	& $5.70\times 10^{1}$ 	& $5.16\times 10^{-2}$	& $1.48\times 10^{-2}$
	 &	\vline  	& $1.28\times 10^{1}$	& $1.18\times 10^{-2}$	& $3.99\times 10^{-3}$ 	\\
Total Bkgd. 		& $6.18\times 10^{1}$ 	& $7.84\times 10^{-2}$	& $2.38\times 10^{-2}$
	&	\vline 	& $1.38\times 10^{1}$	& $1.92\times 10^{-2}$	& $6.61\times 10^{-3}$	\\
Signal - LH 		& - 					& -					& -		
	&	\vline 	& - 					& -					& -		\\
Signal - GM 		& $0.30\times 10^{-1}$ 	& $1.01\times 10^{-2}$ 	& $5.16\times 10^{-3}$
	&	\vline 	& $1.30\times 10^{-2}$	& $4.88\times 10^{-3}$	& $2.56\times 10^{-3}$ 	\\
Signal - LR 		& $1.03\times 10^{-1}$	& $3.56\times 10^{-2}$	& $1.80\times 10^{-2}$
	&	\vline 	& $3.72\times 10^{-2}$	& $1.41\times 10^{-2}$	& $7.44\times 10^{-3}$	\\

\hline
$\bf ZZ$ 			& 					& 					&   
	&	\vline 	& 					& 					& 	\\
\hline
EW - LH 			& $1.59\times 10^{-2}$	& $4.67\times 10^{-3}$	& - 
	&	\vline 	& $5.58\times 10^{-3}$ 	& $1.78\times 10^{-3}$ 	& - 	\\
EW - GM 			& $2.12\times 10^{-2}$ 	& $6.77\times 10^{-3}$ 	& - 
	&	\vline 	& $7.01\times 10^{-3}$ 	& $2.38\times 10^{-3}$	& - 	\\
EW - LR 			& $5.28\times 10^{-2}$ 	& $1.86\times 10^{-2}$	& - 
	&	\vline 	& $1.51\times 10^{-2}$ 	& $5.87\times 10^{-3}$ 	& - 	\\
EW - SM  			& $3.20\times 10^{-3}$ 	& $1.83\times 10^{-4}$ 	& - 
	&	\vline 	& $1.30\times 10^{-3}$ 	& $4.80\times 10^{-5}$ 	& - 	\\
QCD Bkgd.	 	& $8.00\times 10^{-2}$  	& $1.50\times 10^{-3}$ 	& - 
	&	\vline 	& $1.67\times 10^{-2}$ 	& $3.49\times 10^{-4}$	& - 	\\
Total Bkgd. 		& $8.32\times 10^{-2}$ 	& $1.68\times 10^{-3}$ 	& - 
	&	\vline 	& $1.80\times 10^{-2}$ 	& $3.97\times 10^{-4}$ 	& - 	\\
Signal - LH 		& $1.27\times 10^{-2}$ 	& $4.49\times 10^{-3}$ 	& - 
	&	\vline 	& $4.28\times 10^{-3}$ 	& $1.73\times 10^{-3}$ 	& - 	\\
Signal - GM	 	& $1.80\times 10^{-2}$	& $6.59\times 10^{-3}$ 	& - 
	&	\vline 	& $5.71\times 10^{-3}$ 	& $2.33\times 10^{-3}$ 	& - 	\\
Signal - LR   		& $4.96\times 10^{-2}$ 	& $1.84\times 10^{-2}$ 	& - 
	&	\vline 	& $1.38\times 10^{-2}$ 	& $5.82\times 10^{-3}$ 	& - 	\\

\end{tabular}
\end{ruledtabular}
\label{tab:xsect}
\end{center}
\end{table*}

\begin{table*}[t]
\begin{center}
\caption{The integrated luminosity (in fb$^{-1}$) required for a statistical significance of  
$S/\sqrt{B} \geq 5$ and at least 10 signal events, after the cuts of Table \ref{tab:cuts} 
have been imposed.  These results were obtained for a triplet vev of 
$v^\prime = 39$ GeV, using the cross section values listed in Table \ref{tab:xsect}.
\label{tab:lumi}}
\begin{ruledtabular}
\begin{tabular}{lcccccc}
	 & \multicolumn{2}{c}{Littlest Higgs} & \multicolumn{2}{c}{Georgi-Machacek} & \multicolumn{2}{c}{Left-Right Symmetric} \\
	& $M_\Phi=1.0$ TeV & $M_\Phi=1.5$ TeV & $M_\Phi=1.0$ TeV & $M_\Phi=1.5$ TeV & 
	$M_\Phi=1.0$ TeV & $M_\Phi=1.5$ TeV \\
Channel	& ( fb$^{-1}$) & ( fb$^{-1}$) & ( fb$^{-1}$) & ( fb$^{-1}$) & ( fb$^{-1}$) & ( fb$^{-1}$) \\
\hline
$W^{\pm}W^{\pm}$	& 41		& 118	& 41	 	& 119	& 41		& 117	\\
$W^{\pm}Z$ 		& 3300 	& 4350	& 171	& 474	& 591	& 877	\\
$W^+W^-$ 		& - 		& - 		& 22300	& 25200	& 1840	& 2980	\\
$ZZ$ 			& 2230 	& 5780 	& 1520 	& 4290	& 543 	& 1720	\\
\end{tabular}
\end{ruledtabular}
\end{center}
\end{table*}

The observability of a Higgs triplet signal in a $VV$ channel is determined by the relative size of the signal to background, and by the assumed integrated luminosity.  So, for example, in the Littlest Higgs model with $M_{\Phi}=1.0$~TeV and $v'=39$~GeV, the 
cross section after cuts in the $W^\pm W^\pm$ channel is 0.244~fb, compared to the cross section of the combined backgrounds
of 0.0798~fb.  Assuming 100~fb$^{-1}$, this would result in roughly a $9 \sigma$ effect.  One could invert 
this and ask what luminosity would be required to produce a $5\sigma$ discovery (i.e. $S/\sqrt{B} \geq 5$).  
For this example, the luminosity required would be 34~fb$^{-1}$, resulting in $\sim8$ signal events.  
It may be possible to further optimize our cuts in order to enhance the event rate at the expense of lowering the signal to background ratio.  In any case, there is a tradeoff and this would also depend on detector details that are beyond the scope of this paper.
To ensure that a substantial number of signal events are observed, we impose a second criteria that there must be at least 10 signal events, resulting in a required integrated luminosity of 41~fb$^{-1}$.  Following this procedure, in Table \ref{tab:lumi} we give the integrated luminosities required to produce a $5\sigma$ signal with at least 10 signal 
events for the various models for triplet masses of 1.0 and 1.5~TeV.  

The early years of LHC running are 
expected to produce $\sim10$ fb$^{-1}$/year.  It is therefore unlikely that a heavy Higgs triplet would
be observed at the LHC in this period of time.  Once design luminosity is achieved, it is expected that
$\sim100$ fb$^{-1}$ will be produced per year.  In this case it should be possible to observe 
a doubly-charged Higgs triplet in any of these models with the assumed parameter values.  In the $W^{\pm}Z$ 
channel, it seems that only the charged scalar in the Georgi-Machacek model could be observed.  
The $W^{+}W^{-}$ channel is the least promising for observing a Higgs triplet due to the large QCD 
and $t$-quark backgrounds in this channel.

The values in Tables \ref{tab:xsect} and \ref{tab:lumi} were obtained assuming a triplet vev of $v' = 39$ GeV, 
which is the upper bound in the Georgi-Machacek model.  However, in the Littlest Higgs model this parameter is 
constrained to $v' \lesssim 4$~GeV, while in the Left-Right Symmetric model the constraint is 
$v' \lesssim 2$~GeV.
The Higgs triplet production cross section is proportional to $v'^2$, so, with these upper limits, the 
cross sections would be reduced 
by a factor of roughly $10^{-2}$ and $2 \times 10^{-3}$ for the Littlest Higgs and Left-Right Symmetric 
models respectively.  With the resulting cross section values, it is clear that to observe any of the 
TeV-scale Higgs triplet bosons of the 
Littlest Higgs or Left-Right Symmetric models at the LHC using leptonic final states would require 
several $\mathrm{ab}^{-1}$ of integrated luminosity and would await the SLHC.  
However, as is usual in model building, it is possible that the existing constraints could 
be evaded, making these processes more useful than we pessimistically conclude.

\section{Summary}
In this paper, we presented the results of a study of Higgs triplet production via vector 
boson scattering at the LHC. The production cross sections are highly sensitive to the triplet
vacuum expectation value ($v'$) which is already tightly constrained by precision electroweak data.
Of the three models considered, the Georgi-Machacek model can have the largest cross sections due to
the weakest constraint on $v'$.  The $W^\pm W^\pm$ channel is the most promising discovery channel due
to its distinctive final state.  However,  it is still possible to observe 
a signal for the $W^\pm Z$ channel for certain regions of parameter space.  The observation
of this channel is a smoking gun for a Higgs triplet, as this process does not occur in Two Higgs 
Doublet Models.  Discovery of the neutral Higgs triplet state is least promising due to the 
 large QCD and $t$-quark backgrounds in the $W^+ W^-$ channel and the small signal cross
section in the $Z Z \to l^+ l^- l^+ l^-$ channel.  The 
 $Z Z \to l^+ l^- \nu \bar{\nu}$ channel might slightly improve the $\Phi^0$ discovery potential, 
but we have not included this in our analysis and leave this for a future study.  
High luminosity is required for some of the channels and models, which will require the SLHC.

One of the goals of this work was to determine whether scalar triplet production via vector 
boson scattering at the LHC 
could be used to distinguish between models with Higgs triplet bosons.  Our conclusions are 
qualified.  The current constraints on the triplet vev in the Littlest Higgs and 
Left-Right Symmetric models result in production cross sections that would require significant luminosity
to be able to distinguish triplet scalars from Standard Model backgrounds,
higher than the expected LHC luminosities.
In these scenarios, these
measurements could at best constrain the allowed parameter space.  However, if the $W^\pm W^\pm$ 
and $W^\pm Z$ signals are observed, it would be possible to distinguish the three models  
we studied by computing the ratio of the rates in those two channels.  In doing so, the dependence 
on the triplet vev cancels out, and this ratio would be sensitive to the details each model.  
 
We focused our work on the fully leptonic final states, which are easiest to distinguish from Standard Model
backgrounds.   The ``silver-plated'' semileptonic decay modes have significantly higher branching ratios 
and could potentially improve the statistics;
the branching ratios are ${\cal B}(W^\pm W^\pm \to l^\pm \nu l^\pm \nu) \sim 5\%$ versus  
${\cal B}(W^\pm W^\pm \to l^\pm \nu q\bar{q}) \sim 29\%$, 
${\cal B}(W^\pm Z \to l^\pm \nu l^+l^- )\sim 1.5\%$ versus
{\cal B}($W^\pm Z \to l^\pm \nu q\bar{q} + q\bar{q} l^+l^-)\sim 20\%$, and 
${\cal B}(ZZ\to l^+ l^- l^+ l^-)\sim 0.44\%$ versus ${\cal B}(ZZ\to l^+ l^- q\bar{q})\sim 9\%$.
However, much depends on the efficiency in reconstructing $W$ and $Z$ bosons in their hadronic modes.  In 
addition, the kinematic cuts would need to be modified to take into account the hadronic final states.  
We leave this excercise for a future study.

\acknowledgments

The authors gratefully acknowledge P. Kalyniak for her suggestion that led to this work and
H. Logan for her numerous comments and suggestions that greatly improved this paper.  
We also thank H. Hou, D. Morrissey, and J. Reuter for helpful discussions and communications.
This research was supported in part by the Natural Sciences and Engineering Research Council of Canada.



\begin{thebibliography}{99}

\bibitem{Morrissey:2009tf}
  D.~E.~Morrissey, T.~Plehn and T.~M.~P.~Tait,
  arXiv:0912.3259 [hep-ph].

\bibitem{Accomando:2006ga}
  E.~Accomando {\it et al.},
  arXiv:hep-ph/0608079.

\bibitem{Nath:2010zj}
  P.~Nath {\it et al.},
  arXiv:1001.2693 [hep-ph].

\bibitem{ArkaniHamed:2002qy}
  N.~Arkani-Hamed, A.~G.~Cohen, E.~Katz and A.~E.~Nelson,
  JHEP {\bf 0207}, 034 (2002)
  [arXiv:hep-ph/0206021].
  
  
\bibitem{Burdman:2002ns}
  G.~Burdman, M.~Perelstein and A.~Pierce,
  Phys.\ Rev.\ Lett.\  {\bf 90}, 241802 (2003)
  [Erratum-ibid.\  {\bf 92}, 049903 (2004)]
  [arXiv:hep-ph/0212228].

\bibitem{Han:2003wu}
  T.~Han, H.~E.~Logan, B.~McElrath and L.~T.~Wang,
  Phys.\ Rev.\  D {\bf 67}, 095004 (2003)
  [arXiv:hep-ph/0301040].

\bibitem{Han:2005ru}
  T.~Han, H.~E.~Logan and L.~T.~Wang,
  JHEP {\bf 0601}, 099 (2006)
  [arXiv:hep-ph/0506313].
  
\bibitem{Perelstein:2005ka}
  M.~Perelstein,
  Prog.\ Part.\ Nucl.\ Phys.\  {\bf 58}, 247 (2007)
  [arXiv:hep-ph/0512128].

\bibitem{Azuelos:2004dm}
  G.~Azuelos {\it et al.},
  Eur.\ Phys.\ J.\  C {\bf 39S2}, 13 (2005)
  [arXiv:hep-ph/0402037].

\bibitem{Hubisz:2004ft}
  J.~Hubisz and P.~Meade,
  Phys.\ Rev.\  D {\bf 71}, 035016 (2005)
  [arXiv:hep-ph/0411264].


\bibitem{Asakawa:2006gm}
  E.~Asakawa, S.~Kanemura and J.~Kanzaki,
  Phys.\ Rev.\  D {\bf 75}, 075022 (2007)
  [arXiv:hep-ph/0612271].
  
  
\bibitem{Georgi:1985}
  H.~Georgi and M.~Machacek,
  Nucl.\ Phys.\ B {\bf 262}, 463 (1985)

  
\bibitem{Mohapatra:1975}
  R.~N.~Mohapatra and J.~C.~Pati,
  Phys.\ Rev.\ D {\bf 11}, 566 (1975)

\bibitem{Senjanovic:1975rk}
  G.~Senjanovic and R.~N.~Mohapatra,
  Phys.\ Rev.\  D {\bf 12}, 1502 (1975).
  
\bibitem{Frampton:1992wt}
  P.~H.~Frampton,
  Phys.\ Rev.\ Lett.\  {\bf 69}, 2889 (1992).

  
\bibitem{Pisano:1991ee}
  F.~Pisano and V.~Pleitez,
  Phys.\ Rev.\  D {\bf 46}, 410 (1992)
  [arXiv:hep-ph/9206242].

\bibitem{CiezaMontalvo:2006zt}
  J.~E.~Cieza Montalvo, N.~V.~Cortez, J.~Sa Borges and M.~D.~Tonasse,
  Nucl.\ Phys.\  B {\bf 756}, 1 (2006)
  [Erratum-ibid.\  B {\bf 796}, 422 (2008)]
  [arXiv:hep-ph/0606243].
  
  
\bibitem{Csaki:2003dt}
  C.~Csaki, C.~Grojean, H.~Murayama, L.~Pilo and J.~Terning,
  Phys.\ Rev.\  D {\bf 69}, 055006 (2004)
  [arXiv:hep-ph/0305237].

\bibitem{Csaki:2003zu}
  C.~Csaki, C.~Grojean, L.~Pilo and J.~Terning,
  Phys.\ Rev.\ Lett.\  {\bf 92}, 101802 (2004)
  [arXiv:hep-ph/0308038].

\bibitem{Belyaev:2009ve}
  A.~S.~Belyaev, R.~Sekhar Chivukula, N.~D.~Christensen, H.~J.~He, M.~Kurachi, E.~H.~Simmons and M.~Tanabashi,
  arXiv:0907.2662 [hep-ph].

\bibitem{He:2007ge}
  H.~J.~He {\it et al.},
  Phys.\ Rev.\  D {\bf 78}, 031701 (2008)
  [arXiv:0708.2588 [hep-ph]].

\bibitem{Alves:2009aa}
  A.~Alves, O.~J.~P.~Eboli, D.~Goncalves, M.~C.~Gonzalez-Garcia and J.~K.~Mizukoshi,
  arXiv:0907.2915 [hep-ph].

\bibitem{Birkedal:2004au}
  A.~Birkedal, K.~Matchev and M.~Perelstein,
  Phys.\ Rev.\ Lett.\  {\bf 94}, 191803 (2005)
  [arXiv:hep-ph/0412278].
  
\bibitem{Cheung:2008zh}
  K.~Cheung, C.~W.~Chiang and T.~C.~Yuan,
  Phys.\ Rev.\  D {\bf 78}, 051701 (2008)
  [arXiv:0803.2661 [hep-ph]].

\bibitem{Dorsner:2007fy}
  I.~Dorsner and I.~Mocioiu,
  Nucl.\ Phys.\  B {\bf 796}, 123 (2008)
  [arXiv:0708.3332 [hep-ph]].

\bibitem{Han:2005nk}
  T.~Han, H.~E.~Logan, B.~Mukhopadhyaya and R.~Srikanth,
  Phys.\ Rev.\  D {\bf 72}, 053007 (2005)
  [arXiv:hep-ph/0505260].

\bibitem{Gunion:1989ci}
  J.~F.~Gunion, R.~Vega and J.~Wudka,
  Phys.\ Rev.\  D {\bf 42}, 1673 (1990).
  
  
\bibitem{Huitu:1996su}
  K.~Huitu, J.~Maalampi, A.~Pietila and M.~Raidal,
  Nucl.\ Phys.\  B {\bf 487}, 27 (1997)
  [arXiv:hep-ph/9606311].


\bibitem{Gunion:1989in}
  J.~F.~Gunion, J.~Grifols, A.~Mendez, B.~Kayser and F.~I.~Olness,
  Phys.\ Rev.\  D {\bf 40}, 1546 (1989).
  
  
\bibitem{Azuelos:2005uc}
  G.~Azuelos, K.~Benslama and J.~Ferland,
  J.\ Phys.\ G {\bf 32}, 73 (2006)
  [arXiv:hep-ph/0503096].


\bibitem{Barger:1990py}
  V.~D.~Barger, K.~m.~Cheung, T.~Han and R.~J.~N.~Phillips,
  Phys.\ Rev.\  D {\bf 42}, 3052 (1990).


\bibitem{Bagger:1993zf}
  J.~Bagger {\it et al.},
  Phys.\ Rev.\  D {\bf 49}, 1246 (1994)
  [arXiv:hep-ph/9306256].

\bibitem{Bagger:1995mk}
  J.~Bagger {\it et al.},
  Phys.\ Rev.\  D {\bf 52}, 3878 (1995)
  [arXiv:hep-ph/9504426].

\bibitem{Kilgore:1996pj}
  W.~B.~Kilgore,
{\it In the Proceedings of 1996 DPF / DPB Summer Study on New Directions for High-Energy Physics (Snowmass 96), Snowmass, Colorado, 25 Jun - 12
Jul 1996, pp STC132}
  [arXiv:hep-ph/9610375].
  
  


\bibitem{Schmaltz:2005ky}
  M.~Schmaltz and D.~Tucker-Smith,
  Ann.\ Rev.\ Nucl.\ Part.\ Sci.\  {\bf 55}, 229 (2005)
  [arXiv:hep-ph/0502182].

   
\bibitem{ArkaniHamed:2001nc}
  N.~Arkani-Hamed, A.~G.~Cohen and H.~Georgi,
  Phys.\ Lett.\  B {\bf 513}, 232 (2001)
  [arXiv:hep-ph/0105239].
  
\bibitem{Perelstein:2007}
  M.~Perelstein,
  Prog.\ Part.\ Nucl.\ Phys  {\bf 58}, 247 (2007)
  [arXiv:hep-ph/0512128v1].  
  
  

  
\bibitem{Chen:2004}
  M.-C.~Chen and S.~Dawson,
  Phys.\ Rev.\ D {\bf 70}, 015003 (2004)
  [arXiv:hep-ph/0311032v3].
  

\bibitem{Haber:2000}
  H.~Haber and H.~Logan,
  Phys.\ Rev.\ D {\bf 62}, 015011 (2000)





\bibitem{Deshpande:1990ip}
  N.~G.~Deshpande, J.~F.~Gunion, B.~Kayser and F.~I.~Olness,
  Phys.\ Rev.\  D {\bf 44}, 837 (1991).


\bibitem{Amsler:2008zzb}
  C.~Amsler {\it et al.}  [Particle Data Group],
  Phys.\ Lett.\  B {\bf 667}, 1 (2008).

  
\bibitem{Alwall:2007st}
  J.~Alwall {\it et al.},
  JHEP {\bf 0709}, 028 (2007)
  [arXiv:0706.2334 [hep-ph]].


  
\bibitem{Dawson:1985}
  S.~Dawson,
  Nucl.\ Phys.\  B {\bf 249}, 42 (1985).

\bibitem{Cornwall:1974}
  J.~Cornwall, D.~Levin and G.~Tiktopoulos,
  Phys.\ Rev.\  D {\bf 10}, 1145 (1974).


\bibitem{moats}
K. Moats,  M.Sc. thesis Carleton University (2007).


\bibitem{Han:2009em}
  T.~Han, D.~Krohn, L.~T.~Wang and W.~Zhu,
  JHEP {\bf 1003}, 082 (2010)
  [arXiv:0911.3656 [hep-ph]].


\bibitem{Barger:1983jx}
  V.~D.~Barger, A.~D.~Martin and R.~J.~N.~Phillips,
  Phys.\ Lett.\  B {\bf 125}, 339 (1983).

\bibitem{Barger:1987re}
  V.~D.~Barger, T.~Han and J.~Ohnemus,
  Phys.\ Rev.\  D {\bf 37}, 1174 (1988).

\bibitem{Pumplin:2002vw}
  J.~Pumplin {\it et al.}, 
  JHEP {\bf 0207}, 012 (2002).
  
\bibitem{Meade:2007}
  P.~Meade and M.~Reece,
  arXiv:hep-ph/0703031.


\end{thebibliography}
\end{document}